\documentclass[onecolumn, 11pt]{IEEEtran}

\usepackage{multirow} \usepackage{comment} \usepackage{enumitem}

\usepackage{amsmath} \usepackage{amsfonts} \usepackage{latexsym}
\usepackage{amssymb} \usepackage{bbm}

\usepackage{upref} \usepackage{theorem} \usepackage{graphicx}
\usepackage{subfigure} \usepackage{algorithm} \usepackage{algpseudocode}

\newcommand{\remove}[1]{} 
\usepackage[margin=1in]{geometry}
\usepackage{xcolor}
\excludecomment{versiona} \includecomment{versionb}
 \newtheorem{theorem}{Theorem} 
\newtheorem{lemma}{Lemma} 
 
\newtheorem{definition}{Definition} \newtheorem{claim}{Claim}
 
\newtheorem{remark}{Remark} \newtheorem{construction}{Construction}

\begin{document}


\title{Multi-Version Coding - An Information Theoretic Perspective of Consistent Distributed Storage}
\fontfamily{cmr} \selectfont 

\author{\IEEEauthorblockN{Zhiying Wang, Viveck R. Cadambe*} \thanks{*Zhiying Wang is
with the Department of Electrical Engineering and Computer Science, University of California, Irvine, and her email is
zhiying@uci.edu. Viveck R. Cadambe is with Department of Electrical
Engineering, Pennsylvania State University, and his email is
viveck@engr.psu.edu.}  \thanks{This work is published in part, in the Proceedings of the 2014 IEEE International Symposium on Information Theory (ISIT), June 2014 and in the Proceedings of the 2014 IEEE Annual Allerton Conference on Communications, Control and Signal Processing, Oct 2014.}}

\maketitle



\begin{abstract} In applications of distributed storage systems to distributed computing and implementation of key-value stores, the following property, usually referred to as consistency in computer science and engineering, is an important requirement: as the data stored changes, the latest version of the data must be accessible to a client that connects to the storage system. An information theoretic formulation called multi-version coding is introduced in the paper, in order to study storage costs of consistent distributed storage systems. Multi-version coding is characterized by $\nu$ totally ordered versions of a message, and a storage system with $n$ servers. At each server, values corresponding to an arbitrary subset of the $\nu$ versions are received and encoded. For any subset of $c$ servers in the storage system, the value corresponding to the latest common version, or a later version as per the total ordering, among the $c$ servers is required to be decodable. An achievable multi-version code construction via linear coding and a converse result that shows that the construction is approximately tight, are provided. An implication of the converse is that there is an inevitable price, in terms of storage cost, to ensure consistency in distributed storage systems.

\end{abstract}

\section{Introduction} There is an enormous interest in recent times to understand the role of erasure coding in distributed storage systems. In this paper, we formulate a new information theoretic problem, the \emph{multi-version coding} problem, motivated by applications of distributed storage systems to distributed computing and implementation of key-value stores. The multi-version coding problem captures two aspects that are not considered previously in information theoretic studies of distributed storage systems:
\begin{enumerate}[label=\roman*)]
	\item In several applications, the message (data) changes, and the user wants to get the latest version of the message. In computer science literature \cite{Lynch1996}, the notion of obtaining the latest version of the data is known as \emph{consistency}\footnote{ There are several formal models of consistency studied in distributed systems literature (See for example \cite{Lynch1996,  lamport1979make, vogels2008eventually}). In this paper, we use the term consistency to loosely mean that a user wants the most recent version of the data.}
	\item There is an inherent \emph{asynchrony} in storage systems due to the distributed nature of the system. As a consequence, the new version of the message may not arrive at all servers in the system at the same time.
\end{enumerate}
The design of a consistent data storage service over an asynchronous distributed storage system has been studied carefully in distributed computing theory literature \cite{Lynch1996, ABD}, and forms an integral part of several data storage products used in practice, such as Amazon Dynamo \cite{Decandia}, Apache Cassandra \cite{CassandradB}, and CouchDB \cite{CouchDB}. The main objective of the multi-version coding problem is to understand the storage costs of consistent distributed storage systems from an information theoretic perspective. We begin with an informal description of the problem. We discuss the background and motivation of our problem formulation in Section \ref{sec:motivation}. 

\subsection{Informal Problem Description}
\label{sec:intro_problem}

\begin{figure*} \centering
  \includegraphics[width=\textwidth]{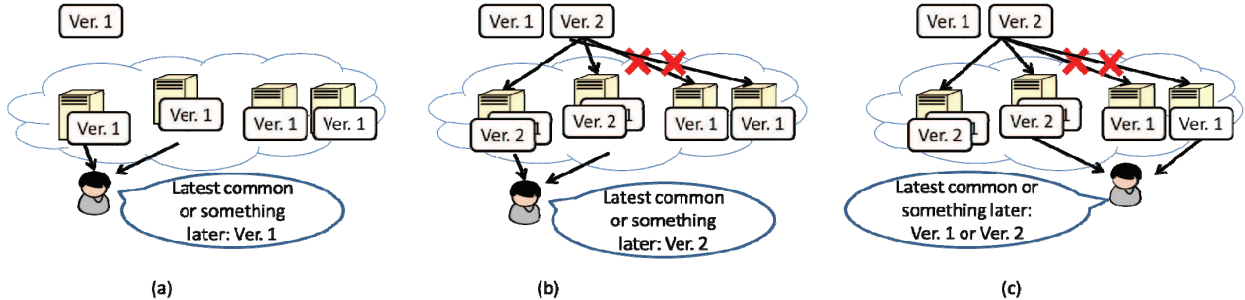}\\ \caption{Storing a file
  with 2 version in $n=4$ nodes. From any $c=2$ nodes, the code should recover
  the latest common version or something later. We denote the old and new
  versions as Version 1 and Version 2 respectively.}\label{fig141005}
  \end{figure*}

Our problem formulation is pictorially depicted in Fig. \ref{fig141005}.
Consider a distributed storage system with a set of $n$ servers. Suppose that
it stores message ${W}_{1}$ using an $n$ length code, such that a decoder can
connect to \emph{any} subset of $c$ servers and decode $W_1$. Suppose an
updated version of the message ${W}_{2}$ enters the system. For reasons that may
be related to network delays or failures, ${W}_{2}$ arrives at some subset of servers,
but not others. We assume that each server is unaware of which servers have
$W_2$ and which do not. The question of interest here is to design a storage
strategy for the servers so that, a decoder can connect to any $c$ servers and
decode the \emph{latest common version} among the $c$ servers, or \emph{some
version later} than the latest common version.  That is, $W_2$ must be
decodable from every set of $c$ servers where each server in the set has
received both $W_1$ and $W_2$. For every set of $c$ servers where there is at
least one server which has not received $W_2$, we require that either $W_1$ or
$W_2$ is decodable.  We intend our storage strategy to be applicable to every
possible message arrival scenario, and every possible subset of servers of size $c$. A possible scenario is depicted in Fig.
\ref{fig141005} for $n=3, c=2$.

Notice that in the storage strategy, a server with both $W_1, W_2$ stores a
function of $W_1$ and $W_2$, whereas a server with only $W_1$ stores a function
of $W_1$. We now describe two simple approaches, \emph{replication} and \emph{simple erasure coding}, that solve this problem.
We assume that the \emph{size} of both versions are equal, that is, the number of bits used to represent $W_1$ is equal to $W_2$. We refer to the size of one version as 
one unit. 
\begin{itemize} \item \emph{Replication:} In this strategy, we assume that each
			server stores the latest version it receives, that is,
			servers with both versions store $W_2$, and servers
			with the first version store $W_1$. Notice that the
			storage cost of this strategy is $1$ unit per server,
			or a total of $n$ units. See Table \ref{tbl4} for an
			example.  \item \emph{Simple Erasure Coding:} In this
				strategy, we use two $(n,c)$ MDS (maximum
				distance separable) codes, one for each version
				separately. A server stores one codeword
				symbol corresponding to every version it
				receives. So, a server with both versions
				stores two codeword symbols resulting in a
				storage cost of $ \frac{2}{c}$ units, whereas,
				a server with only the first version stores
				$\frac{1}{c}$ unit. Notice that for the worst
				case where all servers have both versions,  the
				total storage cost per server is $\frac{2}{c}$
				unit. See Table \ref{tbl5} for an example.
		\end{itemize}
	
We use \emph{worst-case} storage costs to measure the performance of our codes
for simplicity. Therefore, the per server storage cost of replication is equal
to $1$ unit, and that of the simple erasure coding strategy is equal to
$\frac{2}{c}$ units. The singleton bound provides a natural information
theoretic lower bound on the storage cost. In particular, the singleton bound
implies that each node has to store at least $\frac{1}{c}$ units, even for
storing a single version. A natural question of interest is whether we can achieve a storage cost of $\frac{1}{c}$ or whether a new information theoretic lower bound can be found. It is useful to note that asynchrony makes the problem non-trivial. In a synchronous setting, where all the servers receive all the versions at the same time, an MDS code-based strategy where each server stores a codeword symbol corresponding to the latest version received suffices. So, in a synchronous setting, the singleton bound would be tight. 

Our main achievability result provides a code construction, which shows that replication and simple MDS codes are both sub-optimal. It is worth noting that we do not make any assumptions on the correlation between the two versions. Even with our conservative modeling assumption which ignores possible correlation among the versions, we can construct achievable coding schemes that, albeit mildly, improve upon simple erasure coding and replication. 

Our main converse result shows that in the asynchronous setting that we study, the singleton bound is not tight and that the storage cost for any multi-version code cannot be close to $\frac{1}{c}$. Our converse implies that there is an inherent, unavoidable cost of ensuring a consistent storage service because of the asynchrony in the system. 

For the setting described where there are two versions, we provide in this
paper a code construction that achieves a per server storage cost of
$\frac{2}{c+1}$ for odd $c$. When $c$ is even, we achieve a storage cost of
$\frac{2(c+1)}{c(c+2)}$. Table \ref{tbl6} provides an example of our construction
with storage cost of $3/4$ unit, for $n=3,c=2$. Note that our construction outperforms replication
and simple MDS codes. We provide in this paper a converse that shows that the worst
case storage cost cannot be smaller than $\frac{2}{c+1},$ under some mild assumptions. 
The converse implies that our code construction is
essentially optimal for odd values of $c$.

In this paper, we study a generalization of the above problem. In a system with $n$ servers and $\nu$ versions, a
\emph{multi-version code} allows every server to receive any subset of the
$\nu$ versions. Every server encodes according to the versions that it received.
The decoder takes as input, codeword symbols of an arbitrary set of $c$ servers, $c
\le n$, and recovers the latest common version among these servers, or some
version later.  The storage cost is the worst-case storage size per server over
all possible scenarios, that is, over all possible subsets of versions corresponding to
the servers.  In this paper, we provide an information-theoretic
characterization of the storage cost of such codes, including code
constructions and lower bounds for given parameters $n,c,\nu$.

\begin{table} \begin{center} \begin{tabular}{|c|c|c|c|c|} \hline ~  &~ & Server
	1 & Server 2 & Server 3 \\ \hline 
	{Initially Ver. 1 available at all servers} & Ver. 1 & $W_1$ & $W_1$ &
	$W_1$ \\ \hline 
	\multirow{2}{*}{Then, Ver. 2 reaches Servers 1 and 2} & Ver. 1 & ~ & ~ & $W_1$
	\\ ~ & Ver. 2 & $W_2$ & $W_2$ & ~\\ \hline \end{tabular} \end{center}
	\caption{\textnormal{Replication for $n=3,c=2$ with two versions. A server stores $W_1$ if it receives only Version 1; it stores $W_2$ if it receives both versions. Two possible scenarios are shown in the table. Note that the latest version is decodable from every $2$ servers in both scenarios. In general, it can be verified that the latest common version or a later version is decodable from every $c=2$ servers, in every possible scenario.
The storage cost is 1 unit.}} \label{tbl4} \end{table}

\begin{table} \begin{center} \begin{tabular}{|c|c|c|c|c|} \hline ~  &~ & Server
	1 & Server 2 & Server 3 \\ \hline {Initially Ver. 1 available at all servers} & Ver. 1 & $p_1,p_2$ &
	$p_3,p_4$ & $p_1\oplus p_3,p_2\oplus p_4$ \\ \hline
	\multirow{2}{*}{Then, Ver. 2 reaches  Servers 1 and 2} & Ver. 1 & $p_1,p_2$ & $p_3,p_4$ &
	$p_1\oplus p_3,p_2\oplus p_4$ \\ ~ & Ver. 2 & $q_1,q_2$ & $q_3,q_4$ &
	~\\ \hline \end{tabular} \end{center} 
	\caption{\textnormal{Simple erasure coding for $n=3,c=2$ with
		two versions.  Assume each unit is 4 bits, and the bits of the
		two versions are $W_1=(p_1,p_2,p_3,p_4)$,
		$W_2=(q_1,q_2,q_3,q_4)$. Every version is coded with a $(3,2)$
		MDS code where each codeword symbol is a 2-bit vector.
		The 3 codeword symbols for the 3 servers are $(p_1,p_2),(p_3,p_4),(p_1 \oplus p_3, p_2 \oplus p_4)$ for Version 1, and $(q_1,q_2),(q_3,q_4),(q_1 \oplus q_3, q_2 \oplus q_4)$ for Version 2.
A server stores its corresponding codeword symbol of Version 1 if it has only Version 1; it stores codeword symbols of both Version 1 and Version 2 if it receives both versions. Two possible scenarios are shown in the table.  It can be verified that the latest common version is decodable from every $c=2$ servers, in every possible scenario. The storage cost is 4 bits, or equivalently, $1$ unit.}}
\label{tbl5} \vspace{-.6cm} \end{table}

\begin{table} \begin{center} \begin{tabular}{|c|c|c|c|c|} \hline ~  &~ & Server
	1 & Server 2 & Server 3 \\ \hline {Initially Ver. 1 available all servers} & Ver. 1 & $p_1,p_2,p_3$ &
	$p_1,p_2,p_4$ & $p_1,p_2,p_5$ \\ \hline \multirow{2}{*}{Then, Ver. 2 reaches Servers 1 and 2}
	& Ver. 1 & $p_3$ & $p_4$ & $p_1,p_2,p_5$ \\ ~ & Ver. 2 & $q_1,q_2$ &
	$q_3,q_4$ & ~\\ \hline \end{tabular} \end{center}
	\caption{\textnormal{Proposed code for $n=3,c=2$ and
	two versions.  Assume each unit is 4 bits, and the two versions are $W_1=(p_1,p_2,p_3,p_4)$,
		$W_2=(q_1,q_2,q_3,q_4)$. Here $p_5=p_1\oplus p_2\oplus p_3\oplus
	p_4$.  	
Server 1 stores $(p_1,p_2,p_3)$ if it receives only Version 1; it stores $(p_3,q_1,q_2)$ if it receives both versions. Server 2 stores $(p_1,p_2,p_4)$ if it receives only Version 1; it stores $(p_4,q_3,q_4)$ if it receives both versions. Server 3 stores $(p_1,p_2,p_5)$ if it receives only Version 1; it stores $(p_5,q_1 \oplus q_3, q_2 \oplus q_4)$ if it receives both versions. Two possible scenarios are shown in the table. It can be verified that the latest common version or a later version is decodable from every $2$ servers, in every possible scenario. The storage cost is $3$ bits, or equivalently, $3/4$ unit.}}
\label{tbl6} \end{table}

\subsection{Background and Motivation} \label{sec:motivation}

The multi-version coding problem is characterized by two new aspects in its formulation: (i) the idea of consistency in the
decoder, and (ii) the asynchrony in the distributed storage system. We describe some motivating applications and related background literature that inspire our formulation here.

Storing multiple versions of the same message consistently is important in several applications. For instance, the idea of
requiring the latest version of the object is important in \emph{shared memory
systems} \cite{Lynch1996} that form the cornerstone of theory and practice of
multiprocessor programming \cite{Herlihy_Shavit}. In particular, when multiple
threads access the same variable, it is important that the changes made by one
thread to the variable are reflected when another thread reads this variable.
Another natural example comes from key value stores, for instance,
applied to storing data in a stock market, where acquiring the latest stock
value is of significant importance.   

Asynchrony is inherent to the distributed nature of the storage systems used in practice. In
particular, asynchrony occurs due to temporary or permanent failures of
servers, or of transmission between the decoders and the servers. Indeed,
the default model of study in storage systems in the
distributed algorithms literature assumes that communication links can have arbitrarily large delays \cite{Lynch1996}. Since it is more difficult to achieve synchronization in larger systems, asynchrony is an arguably justified modeling choice for distributed storage systems which are expected to scale in practice to cope with rising demands. 

The problem of storing multiple versions of the data consistently in distributed asynchronous storage
systems forms the basis of celebrated results in distributed computing theory \cite{ABD}. From a practical perspective, algorithms designed to ensure
consistency in asynchronous environments form the basis of several commercial storage
products \cite{Decandia, Zookeeper, CouchDB, CassandradB}. We refer the reader to \cite{Decandia} for a detailed description of
the Amazon Dynamo key value store, which describes a replication-based data
storage solution. While \cite{ABD, Decandia} use replication-based
techniques for fault tolerance, the idea of using erasure coding for
consistency has been used in recent distributed computing literature
\cite{Hendricks,Dutta,CadambeCoded_NCA,Dobre}. In fact, these references use
the idea of simple erasure coding that we referred to in Section \ref{sec:intro_problem}. 

{We note that the idea of storing versioned data has acquired some recent interest in information theory literature. In particular, some of the challenges of updating data in distributed storage systems have been studied in \cite{Mazumdar_update, Rouayheb_Synchronizing, Oggier_update, Rawat_update}. These works complement our paper, and their ideas can perhaps be adapted to our framework to build efficient consistent data storage implementations.}

\subsection{Contributions and Organizations} 
Multi-version coding 
provides an information theoretic perspective to the problem of ensuring a consistent data storage service over
an asynchronous distributed storage system. Our problem formulation is geared
towards optimizing the storage cost per server node. We describe the
multi-version coding problem formally in Section \ref{sec2}.  In Section \ref{sec:mainresults}, we formally state our main results: a
multi-version code construction that has a lower cost compared to replication
and naive erasure coding, and an information theoretic lower bound on the
storage cost. The proofs of the main results are provided in Sections \ref{sec:construction}, \ref{sec:converse} and \ref{sec:converse3}. In Section \ref{sec:toy_example}, we demonstrate the utility of multi-version codes using a toy model of asynchronous distributed storage systems.  We discuss related areas of future work in our concluding section, Section \ref{sec:conclude}. 


We describe our achievable multi-version code constructions in Section
\ref{sec:construction}. The construction use a simple linear coding scheme without
coding across versions. Moreover, our code construction satisfies a \emph{causality} property (defined in Section \ref{sec2}) that enables easier implementation, because our encoding strategy is agnostic to the order of arrival of the various message versions at the servers. 

In Section \ref{sec:converse} and \ref{sec:converse3} we prove lower bounds on
the storage cost for $\nu=2$ versions and arbitrary $\nu$, respectively.  Our lower bounds imply that our code constructions are essentially optimal for certain families of parameters, and are close to optimal in general.  
It is worth noting that our problem formulation allows for all possible methods
of encoding the versions. In particular, servers can encode multiple versions
together, and use possibly non-linear methods of encoding the data. The tightness of 
our converse shows that, perhaps surprisingly, encoding each version \emph{separately} using linear codes is close to optimal.

From a technical standpoint, the lower bound argument
is interesting and challenging, especially when the number of versions $\nu$ is larger than $2.$ This is because, in
commonly studied settings in multi-user information theory, the decoder has a
specific set of messages that it wants to recover reliably. In contrast, in multi-version coding,
a decoder is allowed to recover any one of
a subset of messages correctly. As a consequence of the relatively unusual decoding constraint, commonly used methods of deriving
converses need to be modified appropriately to obtain our
lower bound. We provide a more detailed discussion on the
technical aspects of the converse in Section \ref{sec:mainresults}.

{In Section \ref{sec:toy_example}, we describe a toy model of distributed storage that explicitly includes an arrival model for new versions and channels models for the links between the encoders, servers and decoders. We demonstrate the utility of multi-version codes in understanding the storage costs over the described toy model. Our study in Section \ref{sec:toy_example} provides a more refined understanding of the parameters of the multi-version coding problem in terms of the characteristics of a distributed storage system.} {Readers who are interested understanding the applications of multi-version codes, but not the technical details of the construction and the converse, can skip Sections \ref{sec:construction}-\ref{sec:converse3} and read Section \ref{sec:toy_example}. }

\section{System Model: Multi-version Codes}\label{sec2} We begin with some
notations.  For integers $i<j$, we use $[i,j]$ to represent the set
$\{i,i+1,\dots,j\}$.  For integers $i>j$, we define $[i,j]$ as the empty set.
We use $[j]$ to represent the set $[1,j]$. And $[j]$ is an empty set if $j<0$.
For any set of indices $S=\{s_1, s_2, \ldots, s_{|S|}\} \subseteq \mathbb{Z}$
where $s_1 < s_2 < \ldots < s_{|S|}$, and for any ensemble of variables $\{X_i:
i \in S\},$ we denote the tuple $(X_{s_1},X_{s_2},\ldots, X_{s_{|{S}|}})$ by
$X_{S}$.  For a set $\{v_1,\dots,v_n\}$ of elements, we use $v_S$ to denote the
set $\{v_i: i\in S\}$. If $S$ is empty, then $v_S$ is defined to be the empty
set. For sets $S \subseteq T$, we write $T-S$ to be the set difference $\{i:
i\in T, i \notin S\}$.  We use $\log$ to represent log base 2.

We now define the multi-version coding problem. We begin with an informal
definition, and present the formal definition in Definition
\ref{def:storage_code}. The \emph{multi-version coding} problem is
parameterized by positive integers $n,c,\nu,M$ and $q$. We consider a setup
with $n$ servers. Our goal is to store $\nu$ independent versions of the
message, where each version of the message is drawn from the set $[M]$. We
denote the value of the $i$th version of the message by $W_i \in [M]$ for $i
\in [\nu]$.  Each server stores a symbol from $[q]$. Therefore, $\log q$ can be
interpreted as the number of bits stored in a server.  Every server receives an
arbitrary subset of the versions. We denote $\mathbf{S}(i)\subseteq [\nu]$ to
be the set of versions received by the $i$th server. We refer to the set
$\mathbf{S}(i)$ as \emph{the state} of the $i$th server.  We refer to
$\mathbf{S}=(\mathbf{S}(1),\dots,\mathbf{S}(n)) \in \mathcal{P}([\nu])^{n}$ as
the \emph{system state}, where $\mathcal{P}([\nu])$ denotes the power
set of $[\nu]$. 
For the $i$th server, denoting its state $\mathbf{S}(i)$ as
$S=\mathbf{S}(i) =\{s_1, s_2,\ldots, s_{|S|}\}$ where $s_1 < s_2 < \ldots <
s_{|S|},$ the $i$th symbol of the codeword is generated by an encoding function
$\varphi_{S}^{(i)}$ that takes an input, $W_{S}=(W_{s_1}, W_{s_2}, \ldots,
W_{s_{|S|}}),$ and outputs an element in $[q]$.

We assume that there is a total ordering on the versions: if $i < j$, then
$W_j$ is interpreted as a later version of the message as compared with $W_i$.
For any set of servers $T \subseteq [n],$ we refer to $\max \cap_{i \in T}
\mathbf{S}(i)$ as the \emph{latest common version} in the set of servers $T$.
The purpose of multi-version code design is to generate encoding functions such
that, for every subset $T \subseteq [n]$ of $c$ servers, a message $W_m$ should
be decodable from the set $T$, where $m \geq \max \cap_{i \in T} \mathbf{S}(i)$
for every possible system state. The goal of the problem is to find the
smallest possible storage cost per bit stored, or more precisely, to find the
smallest possible value of $\frac{\log q}{\log M}$ over all possible
multi-version codes with parameters $n,c,v,M,q$.

We present a formal definition next.  \begin{definition}[Multi-version code]
\label{def:storage_code} An $(n,c,\nu, M, q)$ \emph{multi-version code}
consists of \begin{itemize} \item encoding functions $$\varphi_{S}^{(i)}:
			[M]^{|S|} \to [q],$$ for every $i \in [n]$ and every $S
		\subseteq [\nu]$, and \item decoding functions  
$$\psi_{\mathbf{S}}^{(T)}: [q]^{c} \to [M] \cup
			\{NULL\},$$ for every set $\mathbf{S} \in
			\mathcal{P}([\nu])^{n}$ and set $T \subseteq [n]$ where
			$|T| = c$, \end{itemize} that satisfy
\small{ \begin{align} &\psi_{\mathbf{S}}^{(T)}
	\left(\varphi_{\mathbf{S}(t_1)}^{(t_1)}({W}_{\mathbf{S}(t_1)}),\dots,\varphi_{\mathbf{S}(t_c)}^{(t_c)}({W}_{\mathbf{S}(t_c)})\right)
	\nonumber\\ =& \begin{cases} {W}_{m}  & \textrm{for some } m
		\geq \max \cap_{i \in T} \mathbf{S}(i),\textrm{if } \cap_{i \in
		T} \mathbf{S}(i) \neq \emptyset, \\ NULL, & \textrm{o.w.,}
	\end{cases}.  \label{eq:decode} \end{align}} for every $W_{[\nu]} \in [M]^{\nu}$, where
	$T=\{t_1, t_2,\ldots, t_c\}$, $t_1< \dots <t_c$.  \end{definition}

 \begin{remark}\label{rmk_decode} Suppose $M \ge \nu$, and let $\mathbf{S}$ be
	the $n$-tuple server state. Consider servers $T \subseteq [n], |T|=c$,
	and the union of their states $S' = \cup_{t \in T} \mathbf{S}(t)$. Then
	for any given tuple $W_{[\nu]}$, the decoding function
	$\psi_{\mathbf{S}}^{(T)}$ decodes either NULL, or a value that is equal
	to $W_j$, for some version $j \in S'$.  \end{remark}

We normalize the storage cost by the size of one version, that is $\log M$.

\begin{definition}[Storage cost of an $(n,c,\nu,M,q)$ multi-version code] The
	\emph{storage cost} of an $(n,c,\nu,M,q)$ multi-version code is defined
	to be equal to $\frac{\log q}{\log M}.$ \end{definition}

As mentioned in the introduction, replication, where the latest version is
stored in every server, i.e.,
$\varphi^{(i)}_{\mathbf{S}_{i}}({W}_{\mathbf{S}_{i}}) =
W_{\max(\mathbf{S}(i))}$ incurs a storage cost of $1$. An alternate strategy
would be to separately encode every version using an MDS code of length $n$ and
dimension $c$, with each server storing an MDS codeword symbol corresponding to
every version that it has received. Such a coding scheme would achieve a
storage cost of $\nu/c$, for sufficiently large $q$.

For parameters $n,c,\nu,$ the goal of the multi-version coding problem is to
find the infimum, taken over the set of all $(n,c,\nu,M,q)$ codes, of the
quantity: $\frac{\log q}{\log M}.$

{It is useful to understand the connection of the parameters of the multi-version coding problem and the physical characteristics of a distributed storage system. The parameter $n$ naturally represents the number of servers across which we intend to encode the data of the storage system. The parameter $c$ is connected to the failure tolerance; in particular, an $(n,c,\nu,M)$ multi-version code can protect against $n-c$ server failures since the latest common version is recoverable among any $c$ nodes.  In Section \ref{sec:toy_example}, we show through a toy model of distributed storage, that the parameter $\nu$ is related to the degree of asynchrony in the system.}

Notice that in our definition, the encoding function of each server 
depends only on the subset of versions that has arrived at
the server, but not on the order of the arrival of the versions. From a practical standpoint, it could be useful to modify the definition of multi-version codes to let the encoding function depend on the order of arrival of the versions. However, in this paper, we use a different approach. We introduce the notion of \emph{causal} multi-version codes that obviates the need for incorporating the order of arrival in the definition.

\begin{definition}[Causal codes] A multi-version code is called \emph{causal}
	if the encoding function satisfies: for all $S \subseteq [\nu], j \in
	S, i \in [n]$, there exists a function $$\hat{\varphi}^{(i)}_{S,j}: [q]
	\times [M] \to [q],$$ such that $$\varphi^{(i)}_{S}(W_S) =
	\hat{\varphi}^{(i)}_{S,j}(\varphi^{(i)}_{S\backslash\{j\}}
	(W_{S\backslash\{j\}}), W_j).$$ \end{definition} To understand the
	notion of causal codes, imagine that a sequence of versions arrive
	at a server in an arbitrary order. If a casual multi-version code
	is used, then the encoding function at the server is only a function of its stored
	information and the value of the arriving version. We anticipate causal
	multi-version codes to be more relevant to practical distributed
	storage systems than non-causal codes. {In fact, we demonstrate the utility of causal multi-version codes in storage systems through our toy model of distributed storage in Section \ref{sec:toy_example}.} All the code constructions that
	we present in this paper are causal.

\section{Main Results} \label{sec:mainresults} In this section, we formally present
the main results of this paper: Theorem \ref{thm:achievability}, which states
the storage cost of an achievable code construction, and Theorem
\ref{thm:converse3}, which states the result of a converse that lower
bounds the storage cost of an arbitrary multi-version code. We present and
discuss Theorem \ref{thm:achievability} in Section \ref{subsec:achievability}. We 
present and discuss Theorem \ref{thm:converse3} in Section \ref{subsec:converse}.

\subsection{Achievability} \label{subsec:achievability}

\begin{theorem} Given parameters $(n,c,\nu)$, there exists a causal
	$(n,c,\nu,M,q)$ multi-version code with a storage cost that is equal to
	$$\max\left\{ \frac{\nu}{c}- \frac{(\nu-1)}{tc}, \frac{1}{t}\right\},$$ where
	$$t =\left\{ \begin{array}{ll} \left
		\lceil\frac{c-1}{\nu}\right\rceil+1, & \text{if } c \geq
		(\nu-1)^2,\\ \left \lceil\frac{c}{\nu-1}\right\rceil, &
		\text{if } c < (\nu-1)^2.\end{array} \right.  $$
		\label{thm:achievability} \end{theorem}

The achievable scheme of Theorem \ref{thm:achievability} has a strictly smaller
storage cost as compared with replication and simple MDS codes.  In particular,  
if $\nu$ is comparable to $c$, our achievable code constructions could improve significantly upon replication and simple MDS codes. If $\nu = c-1$, our storage cost is approximately half the storage cost of the minimum of replication and simple MDS codes for large values of $c$. 
It is instructive to note that if $\nu | (c-1)$, the storage cost is $\frac{\nu}{c+\nu-1}$.

Our code
constructions are quite simple since we do not code across
versions. The main idea of our approach is to carefully allocate the storage
``budget'' of $\log q$ among the various versions in a server's state, and for each version, store an encoded value of the allocated size. 

In \cite{Wang-Cadambe-ISIT}, we studied a special case of the multi-version
code that decodes \emph{only} the latest common version. Here, we allow the
decoder to return a version later than the latest common
version. It is interesting to note that, under the relaxed definition of
multi-version coding presented here, the converse of \cite{Wang-Cadambe-ISIT}
is not applicable. In fact, the achievable scheme of
Theorem \ref{thm:achievability} achieves a storage cost that is lower than the
storage cost lower bound of \cite{Wang-Cadambe-ISIT} by exploiting the fact
that a version that is later than the latest common version can be recovered. We plot the performance of Theorem \ref{thm:achievability} in Fig. \ref{fig141001}.



\subsection{Converse} \label{subsec:converse}

\begin{theorem}\label{thm:converse3} A $(n,c,\nu,M,q)$ multi-version code with
	$n \ge c+\nu-1$ and $M\ge \nu$ must satisfy $$\frac{\log q}{\log M} \ge
	\frac{\nu}{c+\nu-1}-\frac{\log(\nu ^{\nu}
	\binom{c+\nu-1}{\nu})}{(c+\nu-1) \log M }.$$ \end{theorem} In the lower
	bound expression of Theorem \ref{thm:converse3}, the second term on the
	right hand side vanishes as $\log M$ grows. For the case of
	$\nu=2$ versions, we show a somewhat stronger result in section \ref{sec:converse}. In particular, for $\nu=2$, we show that
	the second term in the theorem can be improved to be $\frac{\log
	c}{(c+1)\log M}$.  
	
 The lower bound of Theorem \ref{thm:converse3} indicates that the storage cost,
as a function of $M$, is at least $\nu/(c+\nu-1) + o(1)$.  When $\nu | (c-1)$, the
storage cost of Theorem \ref{thm:achievability} approaches the lower bound of
Theorem \ref{thm:converse3} as $\log M$ grows, and is  therefore asymptotically
optimal. The multi-version coding problem remains open when $\nu \not| (c-1)$. {We establish a connection between the parameter $\nu$ and the degree of asynchrony in a storage system in Section \ref{sec:toy_example}. The converse of Theorem \ref{thm:converse3} in combination with the achievable scheme of Theorem \ref{thm:achievability} therefore implies that the greater the degree of asynchrony in a storage system, the higher the storage cost. In particular, as $\nu$ tends to infinity, the storage cost is one. Therefore, in the limit of infinite asynchrony, the gains of erasure coding vanish, and replication is essentially optimal.  }

The assumption that $\log M$ grows while $c,\nu$ are kept
fixed is a reasonable first order assumption in our study of storage costs because, in systems where storage cost is large, the file size is typically large. The study of multi-version codes for finite $M$ is, nonetheless, an interesting open problem.

In the lower bound proofs, we develop an algorithm that finds
a system state that requires a large storage cost per server to ensure correct decoding. Our approach to deriving the converse has some interesting conceptual aspects.
The standard approach to derive
converses for a noiseless multi-user information theory problem is as follows: (i) express the encoder and decoder constraints using
conditions on the entropy of the symbols, (ii) use Shannon
information inequalities to constrain the region spanned by the entropies of
the variables, and (iii) eliminate the intrinsic variables of the system to get
bounds that must be satisfied by the extrinsic random variables. Usually performing steps (i),(ii) and (iii) requires ingenuity because they tend to be computationally intractable for many problems of interest (see \cite{Tian_433} for example). For the multi-version coding problem, we face some additional challenges since we cannot use steps (i),(ii) and (iii) directly.

To understand the challenges, we re-examine our approach to deriving
a converse in \cite{Wang-Cadambe-ISIT}, where the decoder was restricted to
recovering the latest common version. For the problem in
\cite{Wang-Cadambe-ISIT}, the standard approach to deriving converses
in multi-user information theory was applicable.  In the multi-version coding
problem, note that for state $\mathbf{S}$ we may express the constraint at the encoder as
\begin{equation} \log q \geq
	H(\varphi_{\mathbf{S}(i)}^{(i)}({W}_{[\nu]})), ~~~~H(\varphi_{\mathbf{S}(i)}^{(i)}({W}_{[\nu]})|
	W_{\mathbf{S}(i)}) = 0, \label{eq:entropytwo}\end{equation} 
	for every $i \in [n]$ and every possible state $\mathbf{S}(i) \in \mathcal{P}([\nu])$, and for any distribution on the messages in the system.
	
	In
		\cite{Wang-Cadambe-ISIT}, where we constrained the decoder to
		decode the latest common version, we were able to similarly express the
		constraint at the decoder. For
		example, consider the first $c$ servers and a state $\mathbf{S}$ where the latest common
		version is $k \in [\nu]$ for the servers $[c],$ we expressed
		the decoding constraint as 

\begin{equation}H(W_{k}| \varphi_{\mathbf{S}(1)}^{(1)}({W}_{[\nu]}),
	\varphi_{\mathbf{S}(2)}^{(2)}({W}_{[\nu]}), \ldots,
	\varphi_{\mathbf{S}(c)}^{(c)}({W}_{[\nu]})) = 0
	\label{eq:entropythree}. \end{equation} Note that the above equation can be written for every possible state $\mathbf{S}$. If we assume a
	uniform distribution for all the messages, we have $H(W_{[\nu]}) = \nu
	\log M$. Combining this with (\ref{eq:entropytwo})
	and (\ref{eq:entropythree}), and using Shannon information inequalities, we 
	obtained a bound on $\frac{\log q}{\log M}$.

In the problem we consider here, we can similarly write the constraints
(\ref{eq:entropytwo}). However, in the multi-version
coding problem, the constraint at the decoder cannot be expressed in a manner
analogous to (\ref{eq:entropythree}) because the decoder does not have a
\emph{specific} message to decode. At any given state of the system, a decoder
that connects to $c$ servers is allowed to decode one of several messages. In
particular, imagine that version $k$ is the latest common version for the servers $[c]$, when the system state
is $\mathbf{S}$. Then the decoder is allowed to decode any one of
$W_{k},W_{k+1},\ldots,\ W_{\nu}$ for state $\mathbf{S}$. In fact, one can
conceive of a decoder that may return different message versions for the same
state, depending on the message realization. For instance, one can conceive of
multi-version code where, for a given state $\mathbf{S}$, when the encoded message tuple is $W_{[\nu]} =
(W_1,W_2,\dots,W_{\nu})$, the decoder $\psi_{\mathbf{S}}^{T}$ outputs
$W_k,$ and when the encoded message tuple is $\overline{W}_{[\nu]}=(\overline{W}_1, \overline{W}_2, \ldots,
\overline{W}_{\nu})$, the decoder $\psi_{\mathbf{S}}^{[c]}$ outputs 
$\overline{W}_{k+1}$. As a consequence of the unusual nature of the decoding constraint, the converse proofs in
Sections \ref{sec:converse} and \ref{sec:converse3} has an unusual structure. In particular, we carefully construct some auxilliary variables and write constraints on the entropies of the constructed variables to replace (\ref{eq:entropythree}).
Our approach to deriving converses is potentially useful for understanding \emph{pliable index coding problem} and other recently formulated content-type coding problems \cite{Fragouli_pliable, Fragouli_Content}, where the decoder does not have a unique message, but is satisfied with reliably obtaining one of a given subset of messages.


\section{Code Construction} \label{sec:construction} We describe our
construction in this section. We start with code construction for $\nu=2$
versions, and then generalize the construction for arbitrary $\nu$. In the end,
we show that our construction is a multi-version code in Theorem \ref{thm:3}.

In our construction, each server encodes different versions separately. So that
the total number of bits stored at a server is the sum of the storage costs of
each of the versions in the server state. The encoding strategy at the servers
satisfies the following property: Suppose that Server $i$ is in state $S
\subseteq [\nu]$ and stores $\alpha_{i,v}^{(S)} \log M$ bits of Version $v$,
then Version $v$ can be recovered from the $c$ servers $i_1, i_2, ...i_c$, so
long as \begin{align}\label{eq0427} \sum_{j=1}^{c} \alpha_{i_j,v}^{(S)} \geq 1.
\end{align} Note that such an encoding function can be found for a sufficiently
large value of $q$ using standard coding techniques.  In fact, suppose that the
message $W_v$ is interpreted as a vector over some finite field. We let Server
$i$ store $\alpha_{i,v}^{(S)} \log M$ random linear combinations of elements in
the vector $W_v$. Then Version $v$ can be recovered from any subset of $c$
servers satisfying \eqref{eq0427} with a non-zero probability so long as the
field size is sufficiently large.  As a result, there exists a deterministic
code that decodes Version $v$ if \eqref{eq0427} is satisfied.  We also note
that, in our approach, the storage allocation $\alpha_{i,v}^{(S)}$ only depends
on the server state but not on the server index. Therefore, we can write
$\alpha_{i,v}^{(S)} = \alpha_{v}^{(S)}$ for any Server $i$ at a nonempty state
$S\subseteq [\nu]$.


As a result, to describe our construction, we only need to specify the
parameter $\alpha_{v}^{(S)}$ for every possible server state $S \subseteq
[\nu]$ and every $v \in S$. That is, we only need to specify the information
amount corresponding to Version $v$ stored at a server in state $S$. We denote
$\alpha=\frac{\log q}{\log M}$ as the storage cost. Note that we have $\alpha =
\max_{S \subseteq [\nu]} \sum_{v \in S} \alpha_{v}^{(S)}$.

\begin{definition}[Partition of the servers] For every system state $\mathbf{S}$ where $\mathbf{S}(i) \neq \emptyset$ for all $i \in [n]$,
	we define a partition of the $n$ servers into $\nu$ groups as follows.
	For $i \in [\nu]$, Group $i$ has the set of servers which have Version
	$i$ as the latest version.  \end{definition}

For instance, if $\nu=2$, Group 1 has the servers in state $\{1\}$, and Group 2
contains the servers in states $\{2\}$ and $\{1,2\}$.

\subsection{Code Construction for $\nu=2$} 

We start by describing our construction
for the case of $\nu=2$ versions shown in Table \ref{tbl9121}. In Theorem \ref{thm:cnstr2}, we show that our construction is a multi-version code.

\begin{construction}\label{cnstr0430} Define $$t = \lceil \frac{c-1}{2} \rceil
	+ 1,$$ We construct a code for $\nu=2$ with storage cost
	$\alpha=\frac{2t-1}{tc}$. More specifically, we assign
	$$\alpha_1^{(\{1,2\})} = \alpha-\frac{1}{t}, 
	\alpha_1^{(\{1\})} = \alpha, \alpha_2^{(\{1,2\})} = \alpha_2^{(\{2\})} =
	\frac{1}{t}.$$
\end{construction}

One can see that the code in Table \ref{tbl6} is an example that follows the
above storage allocation. {It is instructive to note that if $c$ is odd, then $\alpha_{1}^{(\{1,2\})}=0,\alpha_{2}^{(\{1,2\})} = \alpha = \frac{2}{c+1}$. This means that if $c$ is odd, each server simply stores $\frac{2}{c+1}\log_{2}M$ bits of the latest version. That is, servers in Group $1$ store $\frac{2}{c+1}\log_{2}M$ bits of Version $1$, and servers in Group $2$ store $\frac{2}{c+1} \log_{2}M$ bits of Version $2$. By the pigeon-hole principle, there are at least $\frac{c+1}{2}$ servers either in Group $1$ or Group $2$; therefore a decoder can connect to any $c$ servers and decode either version $1$ or version $2$. Furthermore, the decoder always obtains the latest common version, or a later version. As a consequence, our storage strategy forms a multi-version code.  We next provide a formal proof next handling odd and even values of $c$ together.}

\begin{theorem}\label{thm:cnstr2} Construction \ref{cnstr0430} is an
	$(n,c,\nu=2,M,q)$ multi-version code with storage cost of
	$\frac{2}{c+1}$ for odd $c$, and $\frac{2(c+1)}{c(c+2)}$ for even $c$.
\end{theorem} \begin{IEEEproof}
%
%
	Consider any set of $c$ servers. We argue that the latest common version or a later version is decodable for every possible state. 

	\noindent \textbf{Case I.} If the latest common version is Version 2, then all the $c$
servers are in Group 2. Since we have $c \ge t$ servers, and each server
contains $1/t$ amount of Version 2, Version 2 is recoverable.\\ \textbf{Case
II.} If the latest common version is Version 1, then  the $c$ servers may be in
state $\{1\}$ or state $\{1,2\}$.  If there are at least $t$ servers in state
$\{1,2\}$, then we can recover Version 2. Otherwise, there are at most $t-1$
servers in state $\{1,2\}$, and at least $c-t+1$ servers in state $\{1\}$. Thus
the total amount of Version 1 in these servers is at least $$(c-t+1)\alpha +
(t-1)(\alpha-1/t) = 1,$$ so we can recover Version 1.  \end{IEEEproof}

\begin{table} \centering \begin{tabular}{|c|c|c|c|} \hline State & $\{1,2\}$
	& $\{1\}$ & $\{2\}$ \\ \hline Ver 1 & $\alpha - 1/t$ &
	$\alpha$ & ~ \\ \hline Ver 2 & $1/t$ & ~ & $1/t$ \\ \hline
\end{tabular} \caption{{\normalfont Storage allocations for code construction
	with $\nu=2$ versions. Note that $t = \lceil \frac{c-1}{2} \rceil + 1$,
	and the storage cost is $\alpha=\frac{2t-1}{tc}$. More specifically,
	$\alpha=1/t$ for odd values of $c$ and $\alpha=\frac{2(c+1)}{c(c+2)}$
for even values of $c$.}\label{tbl9121}} \end{table}

\subsection{Code Construction for an Arbitrary $\nu$}
We generalize our constructions to arbitrary values of $\nu$. We first provide our constructions in \ref{cnstr141001} and then prove in Theorem \ref{thm:3} that our construction is a multi-version code.

\begin{construction}\label{cnstr141001} Define a parameter $t$ as follows.
	\begin{equation} t = \left\{\begin{array}{ll} \lceil \frac{c-1}{\nu}
		\rceil + 1,& c > (\nu-1)^2, \\ \lceil \frac{c}{\nu-1} \rceil, &
		c \leq (\nu-1)^2.\end{array}\right.  \label{eq141005}
	\end{equation} We construct the $(n,c,\nu,M,q)$ code with storage cost
	\begin{align}\label{eq:storagecost} \alpha = \frac{\log q}{\log M} =
		\max\left\{\frac{\nu t -\nu +1}{tc},\frac{1}{t}\right\}.
		\end{align} For state $S$, the parameter $\alpha_{v}^{(S)}$ is
	set as follows: \begin{itemize} \item If Version $j$, $j \ge 2$, is the
				latest version in state $S,$  then
				$\alpha_{j}^{(S)}= \frac{1}{t},$ that is, store
				$\frac{1}{t} \log_{2}M$ bits of Version $j$.
			\item If Version $j$, $j \ge 2$, is the latest version
				in state $S,$ and $\{1\} \in S,$ then,
				$\alpha_{1}^{(S)} = \alpha-\frac{1}{t},$ that
				is, store $\left(\alpha - \frac{1}{t}\right)
				\log_{2}M$ bits of Version $1$.    \item If
					Version 1 is the latest version, namely
					$S=\{1\}$, then
					$\alpha_1^{(S)}=\alpha$. That is, store
					$\alpha \log M$ bits of Version 1.
			\end{itemize} \end{construction}

Note that in our construction, a server in Group $j$ only stores encoded
symbols of Version $j$ and possibly Version $1$.

It is useful to note that if $c > (\nu-1)^2,$ then $t \ge \nu$ and if $c \leq
(\nu-1)^2$, then $t < \nu$.

\begin{remark} It can be readily verified that the storage cost of Construction
	\ref{cnstr141001} can be expressed more explicitly as follows:
	\begin{align*} \alpha = \begin{cases} \frac{1}{t}, & \textrm{if } c \in
		[(t-1)\nu+1,(\nu-1)t], t<\nu, \\ \frac{\nu t -\nu +1}{tc}, &
		\textrm{otherwise.} \end{cases} \end{align*} where $t$ is
		defined as in \eqref{eq141005}.  \end{remark}

\begin{remark} We note that when $\nu|(c-1)$, we have 
$$t=\frac{c+\nu-1}{\nu}$$
	irrespective of whether $c$ is bigger then $(\nu-1)^2$ or not. As a
	result, we have $\alpha=\frac{\nu}{c+\nu-1}$ when $\nu|(c-1)$. {In this case, as per Construction \ref{cnstr141001}, a server in Group $i$ stores $\frac{\nu}{c+\nu-1}\log_{2}M$ bits of version $i$, and does not store any of the older versions. A simple pigeon-hole principle based argument suffices to ensure that any decoder that connects to $c$ servers decodes the latest common version among the servers, or a later version.}
\end{remark}

\begin{figure} \centering
  \includegraphics[width=.5\textwidth]{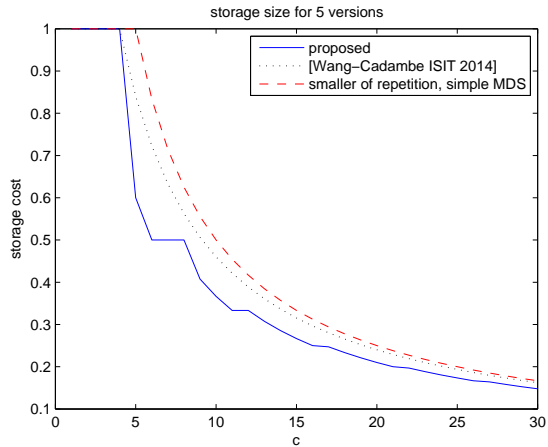}\\
  \caption{Comparison between the construction for the code in Construction
	  \ref{cnstr141001}, the code in \cite{Wang-Cadambe-ISIT}, and the
  smaller of replication and the simple MDS code. We fix $\nu=5$ versions and
  plot results for different number of connected servers,
  $c$.}\label{fig141001} \end{figure}

In Figure \ref{fig141001} we show the storage cost of the construction with
$\nu=5$ versions, we can see the advantage of the proposed code compared to
previous results. 

Table \ref{tbl141005}  is an example with $c=7,\nu=3,\alpha=1/3.$ Notice in
this case, $\nu|(c-1)$, each server only stores information about the latest version it
receives, and does not store any information about any of the older versions.
It is easy to see that when connected to $c=7$ servers with a common version,
at least one version, say Version $i$, can be decoded from $3$ servers in Group
$i$ using similar arguments as the proof of Theorem \ref{thm:cnstr2}.
\begin{table*} \centering \begin{tabular}{|c|c|c|c|c|c|c|c|} \hline ~ &
	\multicolumn{1}{|c|}{Group 1} & \multicolumn{2}{|c|}{Group 2} &
	\multicolumn{4}{|c|}{Group 3}  \\ \hline State & $\{1\}$ &
	$\{2\}$ & $\{1,2\}$ &  $\{3\}$ &  $\{1,3\}$ & 
	$\{2,3\}$ & $\{1,2,3\}$ \\ \hline Version 1 & $1/3$ & ~ & $0$ & ~
	& $0$ & ~ & $0$ \\ \hline Version 2 & ~ & $1/3$ & $1/3$ & ~ & ~ & 0 & 0
	\\ \hline Version 3 & ~ & ~ & ~ & $1/3$ & $1/3$ & $1/3$ & $1/3$  \\
	\hline \end{tabular} \caption{{\normalfont Storage allocations for code
	with $c=7,\nu=3,\alpha=1/3$.}} \label{tbl141005} \end{table*}

\begin{table*} \centering \begin{tabular}{|c|c|c|c|c|c|c|c|} \hline ~ &
	\multicolumn{1}{|c|}{Group 1} & \multicolumn{2}{|c|}{Group 2} &
	\multicolumn{4}{|c|}{Group 3}  \\ \hline State &  $\{1\}$ &
	$\{2\}$ &  $\{1,2\}$ &  $\{3\}$ &  $\{1,3\}$ & 
	$\{2,3\}$ &  $\{1,2,3\}$ \\ \hline Version 1 & $7/15$ & ~ & $2/15$
	& ~ & $2/15$ & ~ & $2/15$ \\ \hline Version 2 & ~ & $1/3$ & $1/3$ & ~ &
	~ & 0 & 0 \\ \hline Version 3 & ~ & ~ & ~ & $1/3$ & $1/3$ & $1/3$ &
	$1/3$  \\ \hline \end{tabular} \caption{{\normalfont Server storage
	allocations for $c=5,\nu=3,\alpha=7/15$.}} \label{tbl141006}
\end{table*} Table \ref{tbl141006} is an example for
$t=3,c=5,\nu=3,\alpha=7/15.$ In this example, the storage cost of the states
are not equal, but one can simply treat the worst-case size as $\alpha$.  One
can check that the above code recovers the latest common version, or a version
that is later than the latest common version.  For example, suppose the latest
common version is Version 1.  \begin{itemize} \item If at least three of the
		$c$ servers are in Group $2$, then Version 2 is recoverable.
	\item If at least three of the $c$ servers are in Group $3,$ then
	Version 3 is decodable.  \item Otherwise, among the $c$ servers, at
		most two servers are in Group $2$, at most two servers are in
		Group $3$, and at least one server is in state $\{1\}$. The
		amount of information of Version 1 in these $c$ servers is at
		least $7/15 + 2/15 \times 4 = 1,$ which implies that Version 1
		is recoverable.  \end{itemize}

\begin{theorem} The code in Construction \ref{cnstr141001} is a casual
	$(n,c,\nu,M,q)$ multi-version code.  \label{thm:3} \end{theorem}
	\begin{IEEEproof} To show a version is recoverable, it suffices to show
		that the total storage allocation for that version in the
		connected servers is at least 1.  Let $j$ be the latest common
		version among $c$ servers. Note that there are at most
		$\nu-j+1$ groups, since Group $1$, Group $2$, \ldots, Group
		$j-1$ are empty.\\ \textbf{Case I.} When $j \ge 2$, there
		exists a group, say Group $k$, with at least $\lceil
		\frac{c}{\nu-j+1} \rceil$ $\ge \lceil \frac{c}{\nu-1} \rceil$
		servers. In our construction, each server in Group $k$ stores
		$\frac{1}{t}$ of Version $k$. To prove the theorem, it suffices
		to show that $\lceil \frac{c}{\nu-1}\rceil \geq t,$ since this
		implies that Version $k$ is recoverable from the servers in
		Group $k$. When $c \le (\nu-1)^2$, then $\lceil \frac{c}{\nu-1}
		\rceil=t$. Therefore, we need to show this for $c > (\nu-1)^2$.
		When $c > (\nu-1)^2$, we have $t = \lceil \frac{c-1}{\nu} \rceil
		+ 1.$  Therefore, we have $c \in [\nu(t-2)+2, \nu(t-1)+1]$.
		Notice also that $t \ge \nu$. These imply the following.
		\begin{align*} & \lceil\frac{c}{\nu-1}\rceil\\ \geq&
			\lceil\frac{\nu(t-2)+2}{\nu-1} \rceil\\ =& \lceil t-1+
			\frac{t-\nu+1}{\nu-1} \rceil \\ \ge& \lceil t-1+
			\frac{1}{\nu-1} \rceil \\ =& t.  \end{align*}
			Therefore, the theorem is proved for the case where $ j
			\geq 2$.\\ \textbf{Case II.} When $j=1$ is the latest
			common version, if Group $i$ has at least $t$ servers
			for any $2\le i \le \nu$, then Version $i$ is
			recoverable and therefore the theorem is proved.
			Otherwise, there are at most $t-1$ servers in Group
			$i$, for all $2\le i \le \nu$, each of which stores
			$\alpha - \frac{1}{t}$ size of Version 1; and thus at
			least $c-(\nu-1)(t-1)$ servers in Group 1, each storing
			$\alpha$ size of Version 1. The total storage cost for
			Version 1 in these servers is at least
			$$(\alpha-\frac{1}{t})(\nu-1)(t-1) +
			\alpha(c-(\nu-1)(t-1)).$$ And by the choice of
			$\alpha$, we know the above amount is at least 1.
			Therefore, Version 1 can be recovered.
    
    If a server is in State $S$, Version $j$ arrives, the server simply need to
    encode based on its stored information and information of Version $j$: (i)
    if $j \le \max S$, the server does nothing; (ii) else the server removes
    $1/t$ amount of information of Version $\max S$, and replace it with $1/t$
    amount of information of Version $j$. Therefore, the construction is
    causal.  \end{IEEEproof}

From the above results, we can prove Theorem \ref{thm:achievability}.
\begin{IEEEproof}[Proof of Theorem \ref{thm:achievability}] Since Construction
	\ref{cnstr141001} is a casual $(n,c,\nu,M,q)$ multi-version code by Theorem
	\ref{thm:3}, and has storage cost as in \eqref{eq:storagecost}, the
	theorem is proved.  \end{IEEEproof}

In fact, the construction in this section is inspired by computer search for
$\nu=3$ and small values of $c$ using integer linear programming on the
allocated storage sizes. 
{In particular, denote by $\mathbf{S}=(\mathbf{S}(1),\dots,\mathbf{S}(c))$ a $c$-tuple server state. Define the latest common version $m(\mathbf{S}) = \max \cap_{i=1}^{c}\mathbf{S}(i)$, for $\cap_{i=1}^{c}\mathbf{S}(i) \neq \emptyset$.
We assume that the versions are coded separately, and use $\alpha_{v}^{(S)}$ to denote the storage allocation of Version $v$ at a server in state $S$, for $v \in S$. Then we have the optimization problem with respect to variables $\alpha, \alpha_{v}^{(S)}$:
\begin{align}
& \textrm{minimize } \alpha, \label{eq:opt}\\
\textrm{s.t. } & \alpha_{v}^{(S)} \ge 0, \textrm{ for all } S \in \mathcal{P}([\nu]), v \in S, \label{eq:constraint0} \\
& \sum_{v \in S} \alpha_{v}^{(S)} \le \alpha, \textrm{ for all } S \in \mathcal{P}([\nu])\label{eq:constraint1}\\
& \bigvee_{v = m(\mathbf{S})}^{\nu} \sum_{i=1}^{c} \alpha_{v}^{(\mathbf{S}(i))}  \ge 1, \textrm{ for all } \mathbf{S} \in \mathcal{P}([\nu])^c, \bigcap_{i=1}^{c}\mathbf{S}(i) \neq \emptyset \label{eq:constraint2}
\end{align}
where $\vee$ is the ``or'' operator. In words, we want to minimize the storage size $\alpha$, subject to the constraint thats the allocation sizes are non-negtive (equation \eqref{eq:constraint0}), every node stores no more than $\alpha$ (equation  \eqref{eq:constraint1}), and the latest common version $m(\mathbf{S})$, or a later version should have enough storage size to ensure recovery (equation \eqref{eq:constraint2}).

We can use the Big $M$ Method \cite{griva2009linear} to convert the ``or'' constraints in \eqref{eq:constraint2} to ``and'' constraints and solve it by integer linear programming. On application of the Big $M$ method, our optimization problem \eqref{eq:opt} can be equivalently expresses as 
\begin{align}
& \textrm{minimize } \alpha, \nonumber\\
\textrm{s.t. } & \alpha_{v}^{(S)} \ge 0, \textrm{ for all } S \in \mathcal{P}([\nu]), v \in S, \nonumber \\
& \sum_{v \in S} \alpha_{v}^{(S)} \le \alpha, \textrm{ for all } S \in \mathcal{P}([\nu])\nonumber\\
& \textrm{ for all } \mathbf{S} \in \mathcal{P}([\nu])^c, \cap_{i=1}^{c}\mathbf{S}(i) \neq \emptyset: \nonumber \\
& \qquad \sum_{i=1}^{c}  \alpha_{v}^{(\mathbf{S}(i))}  \ge 1 - y_v, \forall {v \ge m(\mathbf{S})}, \nonumber \\
& \qquad \sum_{v \ge m(\mathbf{S})}  y_v \le \nu - m(\mathbf{S}) \nonumber \\
& \qquad 0 \le y_v \le 1, y_v \in \mathbb{Z},  \forall {v \ge m(\mathbf{S})}, \nonumber
\end{align}
Plugging in small values of $c$ and $\nu=3$, one can obtain the constructed code as one solution to the above optimization problem.
}

We would like to point out the low complexity to update information in the
servers in our constructions. As the theorem states, our 
constructions are \emph{causal} codes.  Whenever a version arrives that is
the latest among all received ones, the server only needs to delete (a part of)
the older version/s and store the latest version. 
In addition, when $\nu | (c-1)$, no matter how many versions are in the server
state, the server stores information about only the latest version. In this case, every
server only manages a single version and has relatively low complexity compared
to simple MDS coding scheme.

\section{Proof of Converse for 2 Versions} \label{sec:converse} In this
section, we prove Theorem \ref{thm:converse3} for the case of $\nu=2$ versions.
The proof is the inspiration for the proof of general value of $\nu$ in the
next section.

Consider any $(n,c,2,M,q)$ multi-version code, and consider the first $c \le n$
servers. We note here that an arbitrary set of $c$ servers can be considered
for the converse. We consider the first $c$ servers without loss of generality.
In particular, we let the server state be the empty set $\emptyset$ if the server index is
larger than $c$, and we always try to decode from the first $c$ servers.

Informally, the main idea of our argument is as follows. We begin with the
following claim: given the values of the two version, $W_{[2]}=(W_1,W_2)$,
there exist two  system states, $\mathbf{S}_1,$ $\mathbf{S}_{2} \in
\mathcal{P}([\nu])^n$ such that \begin{itemize} \item the states
		$\mathbf{S}_{1},\mathbf{S}_{2}$ differ only in the state of one
	server, say, Server $A$, and \item $W_1$ is decodable from the symbols
		stored among the first $c$ servers in state $\mathbf{S}_{1}$,
		and $W_2$ is decodable from the symbols stored in the first $c$
		servers in state $\mathbf{S}_2$.  \end{itemize}

However, notice that the encoded symbols of the servers $[n]-\{A\}$ are the
same in both states $\mathbf{S}_{1}$ and $\mathbf{S}_{2}$.  This implies that
both Version 1 and Version 2 are decodable from the following $c+1$ symbols:
the $c$ codeword symbols of first $c$ servers in state $\mathbf{S}_{1},$ and
the codeword symbol of the $A$-th server in state $\mathbf{S}_2$. Note that $\mathbf{S}_1$, $\mathbf{S}_2$, and $A$ are chosen based on
the values of $W_{[2]}$, in fact, they may be viewed as functions of $W_{[2]}$.

We now construct $c+1$ variables $Y_{[c]},Z$ as follows: $Y_{i}$ is the value
stored in the $i$-th server for $i \in [c],$ when the server is in state
$\mathbf{S}_1(i)$, and $Z$ is the value stored in the $A$-th server when the
server is in state $\mathbf{S}_2(A)$. Notice that the variables $Y_{i}, i \in [c], Z$ all
belong to $[q]$.


Since these $2$ versions, $W_{1}, W_2$, each of alphabet size $M$, are
decodable from the $c+1$ auxilliary variables $Y_{[c]},Z$ with an alphabet of size $q$, we
need ${(c+1)\log q} \geq {2\log M} + o(1)$. We provide a formal proof next. 
\subsection*{Formal Proof}

Let $\mathcal{S}$ be the set of system states \begin{align*}
	\mathcal{S} =& \{ \mathbf{S} \in \mathcal{P}([\nu])^n : \\
	&\mathbf{S}(i)=\{1,2\}, \forall i \in [x], \\ &\mathbf{S}(i)=\{1\},
	\forall i \in [x+1,c], \\ &\mathbf{S}(i)=\emptyset, \forall i \in
	[c+1,n], \\ &\forall x \in [0,c] \}.  \end{align*} For given values of
	$W_{[2]}$, we define two subsets of $\mathcal{S}$ according to the
	version decoded from Servers $[c]$, denoted by $\mathcal{S}_1$,
	$\mathcal{S}_2$: for $i=1,2$, \begin{align*} \mathcal{S}_i =
		\{\mathbf{S} \in \mathcal{S}: \psi_{\mathbf{S}}^{([c])}(
		\varphi_{\mathbf{S}(1)}^{(1)}(W_{\mathbf{S}(1)}),\dots,\varphi_{\mathbf{S}(c)}^{(c)}(W_{\mathbf{S}(c)})
	)=W_i \}.  \end{align*} 

	We can see that any system
	state in the set $\mathcal{S}$ has the following structure: for some $x \in [0,c],$ the first $x$ servers
	have both versions, servers $[x+1,c]$ have the first version, and the
	remaining servers have no version.  Notice that for any system state in
	$\mathcal{S}$, there exists a latest common version among the first
	$c$ servers.  This means that for every state in $\mathcal{S}$, the
	corresponding decoding function must return Version 1 or Version 2.
	Thus, $\mathcal{S}_1 \cup \mathcal{S}_2 = \mathcal{S}$.  The subset
	$\mathcal{S}_i$ is one where the decoding function returns $W_i,$ the value of
	Version $i$, from the first $c$ servers, for $i=1,2$.  When
	$W_1 \neq W_2$, for any state in $\mathcal{S}$ the decoding function
	returns only one version, therefore $\mathcal{S}_1, \mathcal{S}_2$
	forms a partition of $\mathcal{S}$. When $W_1 = W_2$, for any state we
	can return both versions, so $\mathcal{S}_1=\mathcal{S}_2=\mathcal{S}$.

\begin{claim} For any achievable $(n,c,2,M,q)$ code, and given values
$W_{[2]}$, there are two states $\mathbf{S}_{1}, \mathbf{S}_{2} \in
\mathcal{P}([\nu])^{n}$ such that \begin{itemize} \item The $n$-length tuples
			$\mathbf{S}_{1}$ and $\mathbf{S}_{2}$ differ in one
			element indexed by $A \in [c]$, that is, they differ
			with respect to the state of at most one of the first
			$c$ servers.  \item $\mathbf{S}_{1}  \in \mathcal{S}_{1}$ and $\mathbf{S}_{2}  \in \mathcal{S}_{2}.$
\end{itemize}
				\label{claim1} \end{claim}

\begin{IEEEproof} Assume $W_1=W_2$, then simply take $\mathbf{S}_1$ such that
	their first $c$ elements are all $\{1\}$, and the remaining elements
	are all $\emptyset$. Take $\mathbf{S}_2$ the same as $\mathbf{S}_1$
	except that the first element is $\{1,2\}$. They differ at index $A=1$.
	One can easily check the conditions in the claim.

Assume $W_1 \neq W_2$. Consider a state with the smallest number, $A$, of
occurrences of $\{1,2\}$ in partition $\mathcal{S}_{2}$ and denote this state
as $\mathbf{S}_{2}$. In other words, $$ \mathbf{S}_{2} = \arg \min_{\mathbf{S}
\in \mathcal{S}_{2}} |\{i:\mathbf{S}(i) = \{1,2\}\}|.$$ Let $\mathbf{S}_{1}$ be
a state obtained by replacing the $A$-th element of $\{1,2\}$ of
$\mathbf{S}_{2}$ by $\{1\}$. Notice that, since the number of occurrences of
$\{1,2\}$ in the state tuple $\mathbf{S}_{1}$ is smaller than the number
occurrences of $\mathbf{S}_{2}$, the state $\mathbf{S}_{1}$ does not lie in
partition $\mathcal{S}_{2}.$ Furthermore state $\mathbf{S}_{1}$ lies in $
\mathcal{S}$. Therefore $\mathbf{S}_{1}$ lies in partition $\mathcal{S}_{1}$.
It is easy to verify that states $\mathbf{S}_{1}$ and $\mathbf{S}_{2}$ satisfy
the conditions of the claim.  \end{IEEEproof}

Next, we define $c+1$ variables $Y_{[c]},Z$.  Denote by $A$ the number of
servers in $\{1,2\}$ for $\mathbf{S}_{2}$ found by the proof of Claim
\ref{claim1}, or the largest server index in state $\{1,2\}$ for
$\mathbf{S}_{2}$.
Denoted by $Y_{[c]}$ the values stored in the first $c$ servers when the system
state is $\mathbf{S}_1$: for $i \in [c]$, $$Y_i =
\varphi_{\mathbf{S}_1(i)}^{(i)} (W_{\mathbf{S}_1(i)}). $$ Denote by $Z$ the
value stored in the $A$-th server when the server state is $\mathbf{S}_2(A) =
\{1,2\}$: $$Z = \varphi_{[2]}^{(A)} (W_{[2]}).$$


\begin{IEEEproof}[Proof of Theorem \ref{thm:converse3} for $\nu=2$] Consider
	any $(n,c,\nu=2,M,q)$ code.  Given the value of the variable
	$A$, we can determine the two states $\mathbf{S}_1, \mathbf{S}_2$ as in
	Claim \ref{claim1}.  Therefore, if we are given the values of $A$,
	$Y_{[c]}$ and $Z$, we can determine the values of $W_{[2]}$:
	\begin{align*} W_1 &= \psi_{\mathbf{S}_1}^{([c])}(Y_{[c]}), \\ W_2 &=
		\psi_{\mathbf{S}_2}^{([c])} (Y_{[A-1]},Z,Y_{[A+1,c]}). 
	\end{align*}
	Therefore, there is a bijective mapping from $(Y_{[c]},Z, A)$ to $W_{[2]}$. Therefore, the following equation is true for any distribution over $W_{[2]}$
$$H(W_{[2]}|Y_{[c]},Z,A) = 0.$$ Therefore, $$I(Y_{[c]},Z; W_{[2]}|A) =
H(W_{[2]}|A) = H(W_{[2]}) - I(W_{[2]}; A) \ge H(W_{[2]}) - \log c.$$ The last
inequality holds because the alphabet size of $A$ is at most $c$.  We have the
following chain of inequalities: \begin{align*} &(c+1)\log q  \\ \ge &
	I(Y_{[c]},Z; W_{[2]}|A) \\ \ge & H(W_{[2]}) - \log c.  \end{align*} The
	first inequality follows because $Y_i,Z$ belong to 
	$[q]$ for every $i \in [c]$.  Since the code should work for any distribution of $W_{[2]}$, we assume that $W_1,W_2$ are independent and uniformly distributed over
	$[M]$. Then the theorem statement follows.  \end{IEEEproof}


It is instructive to  observe that in the above proof (and
similarly the proof for general $\nu$ of Theorem \ref{thm:converse3}) that, for different values $W_{[2]}$,
the parameters $A, Y_{[c]},Z$ may take different values. If we constrain the multi-version codes so that the decoding function
$\psi_{\mathbf{S}}^{(T)}$ in Definition \ref{def:storage_code} returns a fixed
version index $m$ given the system state $\mathbf{S}$ and the set of connected
servers $T$, $T \subseteq [n], |T|=c$, then the lower bound can be strengthened \cite{Wang_Cadambe_Allerton}. Our formulation converse proof here is applicable even for multi-version codes where the decoded version index $m$ depends not only on $\mathbf{S}$ and $T$, but could also 
depend on the values $W_{[\nu]}$. 

\section{Proof of Converse for an Arbitrary $\nu$} \label{sec:converse3}
In this section, we provide a proof of Theorem \ref{thm:converse3} for
arbitrary values of $\nu$.  Given an $(n,c,\nu, M,q)$ multi-version code, 
we can obtain a $(c,c,\nu,M,q)$ multi-version code by simply using the
encoding functions corresponding to the first $c$ servers of the given
$(n,c,\nu, M,q)$ code. Furthermore, the storage cost of the $(c,c,\nu,M,q)$
multi-version code is identical to the storage cost of the $(n,c,\nu,M,q)$
multi-version code. Therefore, to derive a lower bound on the storage cost, it
suffices to restrict $n$ to be equal to $c$.  Consider an arbitrary
$(c,c,\nu,M,q)$ multi-version code. Consider the set $\mathcal{W}$ of
message-tuples whose components are distinct, that is, \begin{align}
	\label{eq0506} \mathcal{W} = \{W_{[\nu]}: W_i \neq W_j\textrm{ if } i
\neq j\}.  \end{align} Denote by $\mathbbm{1}_{W{[\nu] \in \mathcal{W}}}$ the
indicator variable: \begin{align*} \mathbbm{1}_{W_{[\nu]} \in \mathcal{W}} =
	\begin{cases} 1, & \textrm{if } W_{[\nu]} \in \mathcal{W}, \\ 0, &
		\textrm{o.w.}.  \end{cases} \end{align*} For a given
		multi-version code, we construct auxilliary variables $Y_{[c-1]},
		Z_{[\nu]}, A_{[\nu]},$ where $Y_{i}, Z_{j} \in [q], i \in
		[c-1], j \in [\nu],$ $1 \le A_1 \le \dots \le A_{\nu} \le c$,
		and a permutation $\Pi:[\nu]\rightarrow [\nu]$, such that there
		is a bijection from values of $\mathcal{W}$ to $(Y_{[c-1]},
		Z_{[\nu]}, A_{[\nu]},\Pi)$. In particular, we describe a mapping $\textbf{AuxVars}$ from
		values in $\mathcal{W}$ to $(Y_{[c-1]}, Z_{[\nu]}, A_{[\nu]},\Pi)$ 
		in Section \ref{secA}, and prove in  
		in Section \ref{secB} that $\textbf{AuxVars}$ is bijective.   

Consider an arbitrary probability distribution on $W_{[\nu]}$, then the
bijection implies that \begin{equation} H(Y_{[c-1]},Z_{[\nu]},A_{[\nu]}, \Pi~|~
	W_{[\nu]}, \mathbbm{1}_{W{[\nu] \in \mathcal{W}}}=1) =  H(W_{[\nu]}~|~
	Y_{[c-1]},Z_{[\nu]},A_{[\nu]}, \Pi, \mathbbm{1}_{W{[\nu] \in
		\mathcal{W}}}=1) = 0 \label{eq:converseequation} \end{equation}
		If we assume a uniform distribution on the elements of
		$W_{[\nu]}$, then the converse of Theorem \ref{thm:converse3}
		follows from the following set of relations.  \begin{align*}
			&\log (q^{c+\nu-1} \binom{c+\nu-1}{\nu} \nu!)\\ & \geq
			H(Y_{[c-1]},Z_{[\nu]}, A_{[\nu]},
			\Pi~|~\mathbbm{1}_{W{[\nu] \in \mathcal{W}}}=1)\\ & =
			I(Y_{[c-1]},Z_{[\nu]}, A_{[\nu]}, \Pi; W_{[\nu]}~|~
			\mathbbm{1}_{W{[\nu] \in \mathcal{W}}}=1)\\ & =
			H(W_{[\nu]}~|~\mathbbm{1}_{W{[\nu] \in
				\mathcal{W}}}=1)\\ & = \log |\mathcal{W}| \\ &
				\geq \log \frac{M^{\nu} \nu!}{\nu^{\nu}},
			\end{align*} where the last inequality follows when $M
			\ge \nu$, and for the first inequality we use the fact
			that $Y_i,Z_j \in [q]$, there are at most $\nu ! $ possibilities for
			$\Pi$, and at most $\binom{c+\nu-1}{\nu}$ possibilities
			of $A_{[\nu]}$.  This implies that
			\begin{align}\label{eq0507} \frac{\log q}{\log M} \geq
				\frac{\nu}{c+\nu-1} - \frac{\nu^{\nu}
				\binom{c+\nu-1}{\nu} }{(c+\nu-1)\log M}
			\end{align} as required.

			To complete the proof, we describe the mapping $\textbf{AuxVars}$ in Algorithm \ref{fig_alg} in Section \ref{secA}, and show in Section \ref{secB} that $\textbf{AuxVars}$ is bijective. Section \ref{sec:properties} describes some useful properties of Algorithm \ref{fig_alg}.

\subsection{Algorithm Description}\label{secA} The function
$\textbf{AuxVars}$ which takes as input, an element $W_{[\nu]}$
from $\mathcal{W}$ and returns variables $Y_{[c-1]},Z_{[\nu]},A_{[\nu]}$ is
described in Algorithm \ref{fig_alg}. The algorithm description involves the
use of a set valued function $\chi$, which we refer to as the \emph{decodable
set function}. We define the function next.

The decodable set function is characterized by the following parameters:
\begin{itemize} \item a positive integer $l \leq c,$ \item a subset $T$ of
		versions, $T \subseteq [\nu];$  \end{itemize} it takes as
	input, \begin{itemize} \item $l$ states ${S}_1,{S}_2,\ldots,{S}_l \in
				\mathcal{P}([\nu]),$ \item and messages
					$W_{[\nu]} \in [M]^{\nu},$
			\end{itemize} and outputs a subset of $[M]$.  Recall that the decoding function $\psi_{\mathbf{S}}^{[c]}$ returns $\textrm{NULL}$ if there is no common version among the servers $[c]$ in state $\mathbf{S}$. The
			decodable set function is denoted as
			$\chi_{l|T}({S}_1,\ldots,{S}_{l},W_{[\nu]})$ and
			defined as 

  \begin{align} &\chi_{l|T}({S}_{1},{S}_{2},\ldots,{S}_{l}, W_{[\nu]})\nonumber
	  \\ =& \bigg\{\psi_{\mathbf{S'}}^{([c])}(X_1,X_2 \ldots, {X}_{c})
	  : \nonumber \\
	  &\mathbf{S'}=({S}_1,\dots,{S}_l,{S'}_{l+1},\dots,{S'}_c), \forall
	  {S'}_j\subseteq T, j \in [l+1,c], \nonumber \\ & X_m =
	  \varphi_{{S}_m}(W_{S_m}), 1 \leq m \leq l   \label{eq0407}\nonumber
	  \\ & X_m = \varphi_{{S'}_m}(W_{S'_m}), l+1 \leq m \leq c \bigg\} - \{\textrm{NULL}\}.
  \end{align} To put it in plain words, the decodable set $\chi_{l|T}$ is the
  set of all non-null values that the decoding function $\psi$ can return,
  given the states of the first $l$ servers, the message realizations of
  $W_{[\nu]},$ when the states of the last $c-l$ servers are restricted to be
  subsets of $T$. It is instructive to note that
  $\chi_{l|T}({S}_{1},{S}_{2},\ldots,{S}_{l}, W_{[\nu]})$ is a subset of
  $\{W_{i}: i \in [\nu]\}$, as stated in the next lemma.

\begin{lemma}\label{lem0501} For every positive integer $l \in [c]$, set $T
	\subseteq [\nu]$ and states  ${S}_1,{S}_2,\ldots,{S}_l \in
	\mathcal{P}([\nu]),$ we have $$\chi_{l|T}({S}_{1},{S}_{2},\ldots,{S}_{l},
	W_{[\nu]}) \subseteq \{W_{i}: i \in [\nu]\}.$$ \end{lemma}
	\begin{IEEEproof} For every collection of $c-l$ states
		$S'_{l+1},S'_{l+2},\ldots, S'_{c} \subseteq T$ such that the
		state $$\mathbf{S}' = ({S}_1,{S}_2,\ldots,{S}_l,
		S'_{l+1},S'_{l+2},\ldots, S'_{c})$$ has a common version, the
		decoding function $\psi_{\mathbf{S}'}$ returns a message value
		that was encoded. Since the encoded message is $W_{[\nu]}$, the
		decoding function returns an element in $\{W_{i}: i \in
	[\nu]\}$. 

If there is no collection of $c-l$ states $S'_{l+1},S'_{l+2},\ldots, S'_{c}
\subseteq T$ such that the state $$\mathbf{S}'= ({S}_1,{S}_2,\ldots,{S}_l,
S'_{l+1},S'_{l+2},\ldots, S'_{c})$$ has a common version, the decoding function
$\psi_{\mathbf{S}^{'}}$ returns $\textrm{NULL}$. In this case, the decodable
set function returns an empty set, which is a subset of $\{W_{i}: i \in
[\nu]\}.$ Therefore $\chi_{l|T}({S}_{1},{S}_{2},\ldots,{S}_{l}, W_{[\nu]})$ is
always a subset of $\{W_{i}: i \in [\nu]\}$.  \end{IEEEproof}

The following property is useful in our description of Algorithm \ref{fig_alg}.
\begin{lemma} Consider messages $W_{[\nu]}$ that have unique values, that is,
	$W_{[\nu]} \in \mathcal{W}$. Then, for any element $W \in
	\chi_{l|T}({S}_{1},{S}_{2},\ldots,{S}_{l}, W_{[\nu]}),$ there is a
	unique positive integer $m \in [\nu]$ such that $W_{m} = W$.
	\label{lem:usefulinalgo} \end{lemma} \begin{IEEEproof} The lemma
		readily follows from noting that there is a one-to-one
		correspondence between $[\nu]$ and $W_{[\nu]}$, and that every
		element $W$ in $\chi_{l|T}$ is also an element in $W_{[\nu]}$
		by Lemma \ref{lem0501}.  \end{IEEEproof}

The decodable set function has an intuitive interpretation when $W_{[\nu]}$ has
unique values, that is, when $W_{[\nu]}\in \mathcal{W}$. If
$\chi_{l|T}({S}_{1},{S}_{2},\ldots,{S}_{l}, W_{[\nu]}) - \{W_{i}: i \in T\}$ is
non-empty, then, loosely speaking, this implies that the first $l$ servers
contain enough information for at least one message in $[\nu]-T$. This is
because the decodable set function  restricts the state of the last $c-l$
servers to be from $T$; as a consequence, if it returns a value corresponding
to a version in $[\nu]-T$, then the first $l$ servers must contain sufficient
information of this version.

\begin{algorithm} \begin{algorithmic}[1] \State
	\textbf{AuxVars}($W_{[\nu]}$) \State Initialize
	$\textrm{VerCount}\gets 1$ \State Initialize $\textrm{ServCount} \gets
	1$	  \State Initialize set $\textrm{VersionsEncountered} \gets
	\{\}$
	  \State Initialize 
	  $Y_{j} \gets 1, Z_{k} \gets 1,A_{k} \gets 1, j \in [c], k \in [\nu]$. 
	  
    \While{$\textrm{VerCount} \le \nu$ and $\textrm{ServCount} \leq c$} \State
    $\mathbf{S}(\textrm{ServCount}) \gets [\nu]- \textrm{VersionsEncountered}$
    \label{l8}          \State $T \gets [\nu]-\textrm{VersionsEncountered}.$
    \label{lT}
	\State $U \gets \left\{W_u: W_u \in
		\chi_{\textrm{ServCount}|T-\{u\}}\left(\mathbf{S}(1),\mathbf{S}(2),\ldots,\mathbf{S}(\textrm{ServCount}),
		W_{[\nu]}\right), u \in T \right\}$ \label{l5}
\If{ $U  \neq \emptyset$ } \label{l6}
\State {$W \gets \max U$} \Comment Natural ordering on $[M]$ for max \label{lW}
\State Let $v \in [\nu]$ such that $W_{v} = W$. \Comment{From Lemma
	\ref{lem:usefulinalgo} there exists a unique $v$} \label{l13} \State
	$A_{\textrm{VerCount}} \gets \textrm{ServCount}$ \label{lA}
	\State $Z_{\textrm{VerCount}}\gets
	\varphi_{\mathbf{S}(\textrm{ServCount})}^{(\textrm{ServCount})}(W_{\mathbf{S}(\textrm{ServCount})})$ \label{lZ}
	  \State $\Pi(\textrm{VerCount}) \gets v$ \label{lpi} \State
	  $\textrm{VersionsEncountered} \gets \textrm{VersionsEncountered} \cup
	  \{v\}$ \State $\textrm{VerCount} \gets \textrm{VerCount}+1$
	  \Else \State $Y_{\textrm{ServCount}} \gets
	  \varphi_{\mathbf{S}(\textrm{ServCount})}^{(\textrm{ServCount})}(W_{\mathbf{S}(\textrm{ServCount})})$\label{lY}
	  \State $ \textrm{ServCount} \gets \textrm{ServCount}+1$ 
	  \EndIf \EndWhile 
	  \State If $\Pi$ is not a permutation of $[\nu]$, set $\Pi$ to be an arbitrary permutation. \Comment As a consequence of Lemma \ref{lem0429} and \ref{rmk8}, this line is never executed.
	  \State Return $Y_{[c-1]},Z_{[\nu]},A_{[\nu]}, \Pi$
  \end{algorithmic} \caption{Function \textbf{AuxVars}: Takes
  input $W_{[\nu]} \in \mathcal{W},$ and outputs variables
  $Y_{[c-1]},Z_{[\nu]},A_{[\nu]},\Pi$.} \label{fig_alg} \end{algorithm}

\begin{figure} \centering
  \includegraphics[width=.5\textwidth]{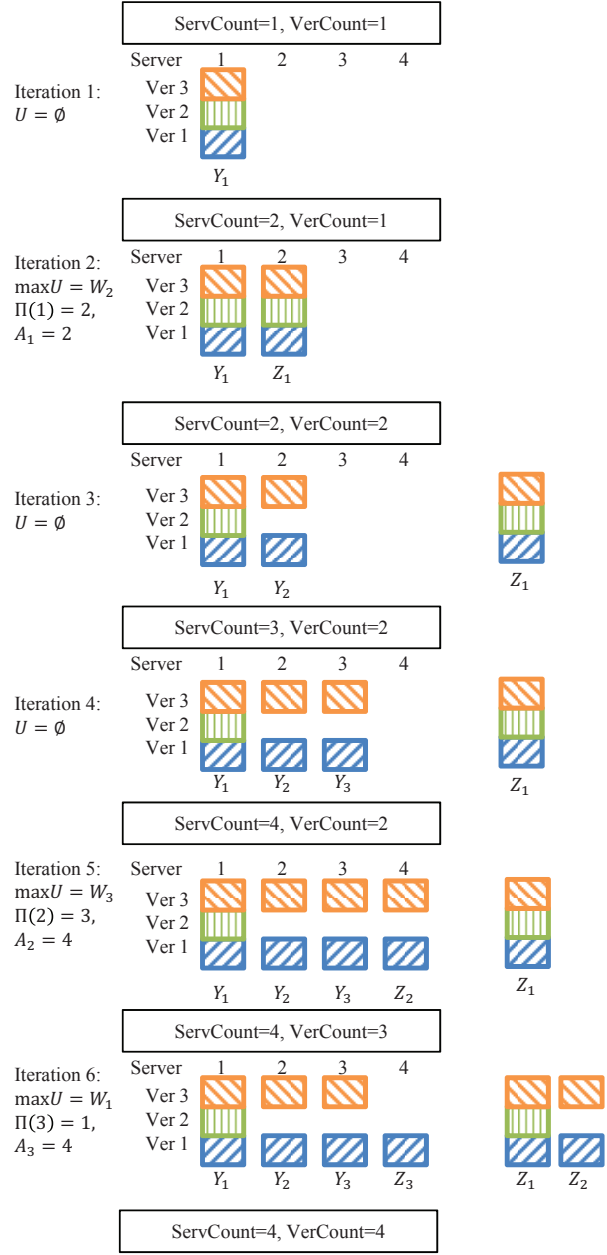}\\ \caption{Example of
	  the algorithm. $c=4,\nu=3$. The resulting server indices are
	  $A_{[3]}=(2,4,4)$, and the permutation on the versions is
  $\Pi=(2,3,1)$.}\label{fig:converse} \end{figure} In Algorithm \ref{fig_alg},
  we describe the function $\textbf{AuxVars}$ that takes as
  input $W_{[\nu]} \in \mathcal{W}$ and returns
  $(Y_{[c-1]},A_{[\nu]},Z_{[\nu]}, \Pi)$. Here, we informally describe the
  algorithm and examine some properties.

In every iteration of the while loop of Algorithm \ref{fig_alg}, either
$\textrm{VerCount}$ increases by $1$, or $\textrm{ServCount}$ increases by $1.$
In particular, if Line \ref{l6} returns true, then $\textrm{VerCount}$
increases by $1$, otherwise $\textrm{ServCount}$ increases by $1$. Therefore,
the while loop terminates, and as a consequence, the algorithm terminates. In
our subsequent discussions, we identify an iteration of the while loop by its
unique $\textrm{VerCount}$--$\textrm{ServCount}$ pair at the beginning of the iteration.

Every iteration of the while loop begins by setting the server state $\mathbf{S}(\textrm{ServCount})$ in Line
\ref{l8}. If Line \ref{l6} is false, then the iteration sets
$Y_{\textrm{ServCount}}$ in line \ref{lY} and then increments $\textrm{ServCount}$. If Line
\ref{l6} is true, then the iteration sets $A_{\textrm{VerCount}},
Z_{\textrm{VerCount}}$ and $\Pi(\textrm{VerCount})$ respectively in Lines \ref{lA},\ref{lZ},\ref{lpi}, and then increments
$\textrm{VerCount}$. In particular, $A_{\textrm{VerCount}}$ is set to the server index $\textrm{ServCount}$, and $\Pi(\textrm{VerCount})$ is set to the version index $\textrm{VerCount}$. Note that $A_{1}$ is the smallest value of $\textrm{ServCount}$ 
such that Line \ref{l6} returns true, that is, it is the smallest integer such
that $\chi_{A_1|[\nu]-\{u\}}([\nu],[\nu],\ldots,[\nu],W_{[\nu]})$ contains
$W_{u}$ for some $u \in [\nu]$. Intuitively speaking, $A_1$ is the smallest integer such that the first $A_1$ servers have enough information about some version $v \in [\nu],$ when the states of these servers are all set to $[\nu]$. {If more than one version in $[\nu]$ returns true for the iteration with $\textrm{ServCount}=A_1$, then $v$ is picked to be the version index corresponding to the maximum value of $\{W_{u}:\chi_{A_1|[\nu]-\{u\}}([\nu],[\nu],\ldots,[\nu],W_{[\nu]})\textrm{ contains }W_u\}.$ The iteration sets $\Pi(\textrm{VerCount})$ to the version index $v$.}

In Figure \ref{fig:converse}, we show an example of a
possible execution of the algorithm for $\nu=3, c=4$ that happens to halt at
$\textrm{ServCount}$ $A_{\nu}=4$ for a particular multi-version code and message tuple
$W_{[3]}$. The states of the servers are set one by one to  $\{1,2,3\}$ as in
Line \ref{l8}. The algorithm proceeds incrementing $\textrm{ServCount}$ in
every iteration where Line \ref{l6} returns false. In an iteration where Line
\ref{l6} returns true, $\textrm{VerCount}$ is incremented. Suppose at $\textrm{ServCount}$
$=2$, Line \ref{l6} returns true for the first time in the execution, and
suppose that $v=2$ in Line \ref{l13}; the algorithm sets
$\Pi(1)=2,A_1=\textrm{ServCount}=2$, and, in the next iteration, the state of
the second server is reset to $[\nu]-\{\Pi(1)\}=\{1,3\}.$ 
Then the algorithm proceeds incrementing $\textrm{ServCount}$ every time Line \ref{l6}
returns false, setting the state of the corresponding server to $\{1,3\}$. Now,
suppose that Line \ref{l6} returns true at $\textrm{ServCount}$ $= 4$, and that $v = 3$ in
Line \ref{l13}. Then $\Pi(2) = \textrm{VerCount} = 3, A_2 = \textrm{ServCount}
= 4$. In the next iteration, the state of server $4$ is set to $\{1\}$.  Then
$A_3$ is set similarly.

{We later show that from the variables $Y_{[3]},Z_{[3]},A_{[3]},\Pi$, one can recover all of the 3 versions $W_{[3]}$. Here we provide an informal overview of the argument. From Iteration 6 in Fig. \ref{fig:converse}, we can observe that $W_{1}$ is equal to $\psi^{[4]}(Y_1, Y_2, Y_3, Z_3),$ where the decoding function $\psi^{[4]}$ is evaluated with states $S_{1}= \{1,2,3\}, S_{2} = S_{3} =\{1,3\}, S_{4}=\{1\}$. The states $S_1, \ldots, S_4$ can be inferred from $A_{[3]}$ and $\Pi$. Similarly, from Iteration 5 in Fig. \ref{fig:converse}, we can observe that $W_{3}$ is equal to $\psi(Y_1, Y_2, Y_3, Z_2)$ with states $S_{1}= \{1,2,3\}, S_{2} = S_{3} =\{1,3\}, S_{4}=\{1,3\}.$ In the converse proof, we will show that, given $W_1, W_3,$ the value $W_2$ can be recovered by using the conditional decoding function as the maximum value of the set $\chi_{2|\{2,3\}}( \{1,2,3\},\{1,2,3\}, W_1)-\{W_1, W_3\},$ which can be evaluated using the values $W_1, W_3, Y_1, Y_2,Z_1$ as 
\begin{align*} &\bigg\{\psi_{\mathbf{S'}}^{([4])}(X_1,X_2, X_3, {X}_{4})
	  : \nonumber \\
	  &\mathbf{S'}=(\{1,2,3\}, \{1,2,3\},{S'}_{3},{S'}_4), \forall
	  {S'}_j\subseteq \{1,3\}, j \in [3,4], \nonumber \\ & X_1 =
	  Y_1, X_2 = Z_1, \nonumber
	  \\ & X_m = \varphi_{{S'}_m}^{(m)}(W_{S'_m}), m=3,4 \bigg\} - \{W_1, W_3, \textrm{NULL}\}.
  \end{align*}
In particular, we will show that the above set is a singleton set, with the element being $W_{2}$.

\subsection{Properties of Algorithm \ref{fig_alg}}
 We next list some useful and instructive {properties of Algorithm \ref{fig_alg}} before proceeding to formally prove that $\textbf{ConverseAuxilliaryVars}$ is invertible.}
\label{sec:properties}
\begin{enumerate}[leftmargin=*,itemindent=*, label= Property (\arabic*)] \item \label{rmk10} If the last iteration of the while loop begins with $\textrm{VerCount}$ $=\nu$,
			the server indices satisfy $1 \le A_1 \le \dots \le A_{\nu} \le c.$ Moreover, for any $t \le \nu$, every iteration of the while loop with
			$\textrm{ServCount} = A_t -1$ has $\textrm{VerCount}$ $\le t$.
 Later, in Lemma \ref{lem0429}, we show that in every execution of Algorithm \ref{fig_alg}, the last iteration of the while loop indeed begins with $\textrm{VerCount} = \nu$. 		
 \item \label{rmk7} For the iteration of the while loop with $\textrm{VerCount}$ and $\textrm{ServCount}$,
			the server states are set by the algorithm to be
			\begin{equation} \mathbf{S}(i)=\begin{cases} ~[\nu], & i
				\in [1,A_{1}-1], A_1>1,\\ 
				~[\nu]-\{\Pi(x): x \in [m]\},
			& i \in [A_m,A_{m+1}-1], m \in [\textrm{VerCount}-2], A_{m+1}>A_m, \\
		~[\nu]-\{\Pi(x): x \in [\textrm{VerCount}-1]\}, & i \in
	[A_{\textrm{VerCount}-1},\textrm{ServCount}].  \end{cases}
	\end{equation} \item \label{rmk11} For the iteration of the while loop with $\textrm{VerCount}$ $=j$, $j
		\ge 2$, and $\textrm{ServCount}$ $=k$, let $\mathbf{S}(i)$ be the state of
		Server $i$, $i \in [k]$.  Let $t < j$.  Consider the last
		iteration with $\textrm{ServCount}$ $=A_t-1$. Suppose that, in this
		iteration, $\textrm{VerCount}$ $=x$, and $\hat{\mathbf{S}}(i)$ is the
		state of Server $i$, $i \in [A_t-1]$.  Then for $i \in
	[A_t-1]$, the states $\mathbf{S}(i)$ are the same as the sates
$\hat{\mathbf{S}}(i)$.  \item \label{rmk5} At the beginning of an iteration of the while loop, the set $\textrm{VersionsEncountered}$ is
	$\{\Pi(1),\dots,$ $\Pi(\textrm{VerCount-1})\}.$ The set $T$ indicates the set of versions not
	encountered, which is $[\nu]-$ $\{\Pi(1),\dots,$ $\Pi(\textrm{VerCount-1})\}$.
\item \label{rmk8} In
		Line \ref{l5}, $U \subseteq \{W_i, i\in T\}$, where
		$T=[\nu]-\textrm{VersionsEncountered}$. Therefore, when Line \ref{l6} returns true, $U \cap \{W_{i}:i \in
		\textrm{VersionsEncountered}\} = \emptyset$. As a consequence,
		$\Pi(\textrm{VerCount}) \notin \{\Pi(1),\dots,
			\Pi(\textrm{VerCount}-1)\}$. Therefore, if the last iteration of the while loop begins with $\textrm{VerCount}=\nu$ and Line \ref{l6} returns true in this iteration, then $\Pi$ is indeed a permutation at the end of the iteration. 
		\item \label{rmkYZ} {Consider the last iteration of the while
				loop of the algorithm where Line \ref{l6}
				returns true. Suppose this iteration begins
				with $\textrm{VerCount}$ $=j$, $\textrm{ServCount}$ $=A_j$. Then we
				have for all $i \in [A_j-1]$, $$Y_{i} =
				\varphi_{\mathbf{S}(i)}^{(i)}(W_{\mathbf{S}(i)}),$$
				where $\mathbf{S}(i)$ is specified in
				\ref{rmk7} with $\textrm{VerCount}$ $=j$, $\textrm{ServCount}$
				$=A_j$. Moreover, for all $i \in [j]$, $$Z_{i}
				=
				\varphi_{\mathbf{S'}(A_i)}^{(A_i)}(W_{\mathbf{S'}(A_i)}),$$
			where $\mathbf{S'}(A_i) = \{\Pi(i),\dots,\Pi(\nu)\}$.
	} \item \label{rmk9} Note that when Line \ref{l6} returns true, even
		though $W_{[\nu]}$ is an input to the function
		$\chi_{\textrm{ServCount}|T-\{u\}}$ in Line \ref{l5}, {for
		every $u \in T$,} we can generate the output of the function
		only from $\mathbf{S}(1),\dots,$ $\mathbf{S}(\textrm{ServCount})$,
		$\{W_{i}: i \in T-\{u\}\},$
		$Y_m=\varphi_{\mathbf{S}(m)}(W_{\mathbf{S}(m)}), 1 \leq m \leq
		\textrm{ServCount}-1$, and $Z_{\textrm{ServCount}} =$
		$\varphi_{\mathbf{S}(\textrm{ServCount})}(W_{\mathbf{S}(\textrm{ServCount})})$. In particular,
		\begin{align} &
			\chi_{\textrm{ServCount}|T-\{u\}}\left(\mathbf{S}(1),\mathbf{S}(2),\ldots,\mathbf{S}(\textrm{ServCount}),
			W_{[\nu]}\right) \nonumber\\ &=
			\bigg\{\psi_{\mathbf{S'}}^{([c])}(X_1,X_2 \ldots,
			{X}_{c}) \neq \textrm{NULL}: \nonumber \\
			&\mathbf{S'}=(\mathbf{S}(1),\ldots,\mathbf{S}(\textrm{ServCount}),
			{S'}_{l+1},\dots,{S'}_c), \forall {S'}_j\subseteq
			T-\{u\}, j \in [l+1,c], \nonumber \\ & X_m = Y_m, 1
			\leq m \leq \textrm{ServCount}-1  \nonumber \\ & X_m =
			Z_m, m = \textrm{ServCount}  \nonumber \\ & X_m =
			\varphi_{{S'}_m}^{(m)}(W_{S'_m}), \textrm{ServCount}+1 \leq m
		\leq c \bigg\}.   \label{eq:rmk9} \end{align} We use this
		property in showing that $\textbf{AuxVars}$ is
		one-to-one.  \end{enumerate}

		We next state Lemmas \ref{lem0428} and \ref{lem0429}. 
Statement (i) of Lemma \ref{lem0428} is useful in proving Lemma \ref{lem0429}. Statement (ii) of Lemma \ref{lem0428} is useful in the proof of Theorem \ref{thm:converse3}, in particular, in inverting $\textbf{AuxVars}$ to obtain $W_{[\nu]}$ from $Y_{[c-1]},Z_{[\nu]},A_{[\nu]}, \Pi$.		
		Lemma \ref{lem0429} shows that at the beginning of the last iteration in the algorithm, the
variable $\textrm{VerCount}$ is equal to $\nu$. This means that all $\nu$
versions returns true at Line \ref{l6} in some iteration of the while loop.


\begin{lemma}\label{lem0428} { (i) Consider any execution of
		$\textbf{AuxVars}$ and consider an iteration of
		the while loop. After Line \ref{lT} is executed in the while
		loop, the following statement is true for any  $u \in T$:
		\begin{align}\label{eq0428}
			\chi_{\textrm{ServCount}|T-\{u\}}\left(\mathbf{S}(1),\mathbf{S}(2),\ldots,\mathbf{S}(\textrm{ServCount}),
			W_{[\nu]}\right) \subseteq \{W_{i}, i \in T\}.
		\end{align} where $\textrm{ServCount}$ represents the value at
		the beginning of the while loop iteration.

(ii) Consider any execution of \textbf{AuxVars} where the final
iteration of the while loop begins with $\textrm{VerCount}=k$ . For any $t \in
[k]$, we have $$ \{W_{\Pi(t)}\}  = \chi_{A_{t} | T - \Pi(t)} ( \mathbf{S}(1),
\ldots \mathbf{S}(A_t), W_{[\nu]}) - \{W_i: i \in T-\{\Pi(t)\} \},$$ where, $T =
\{\Pi(t), \Pi(t+1), \ldots, \Pi(\nu)\}$, and $\mathbf{S}$ is defined as
\ref{rmk7} with $\textrm{VerCount}$ $=t$, $\textrm{ServCount}$ = $A_t$.  } \end{lemma}

\begin{IEEEproof}  (i)  Note that $T=[\nu] -\textrm{VersionsEncountered}$, and
	$\textrm{VersionsEncountered}=$
	$\{\Pi(1),\Pi(2),$ $\ldots,\Pi(\textrm{VerCount}-1)\}$ by \ref{rmk5}.  Notice that when
	$\textrm{VerCount}$ $=1$, the claim is satisfied automatically because $T=[\nu]$.
  
  We prove by contradiction.  We suppose at $\textrm{ServCount}$ $=k$, $k \in [c]$, and
  $\textrm{VerCount}$ $=j$, $j \in [2,\nu]$, equation \eqref{eq0428} is violated, and
  \begin{equation} W_{\Pi(t)} \in
	  \chi_{k|T-\{u\}}\left(\mathbf{S}(1),\mathbf{S}(2),\ldots,\mathbf{S}(j),
	  W_{[\nu]}\right), \label{eq:contradictthis} \end{equation} for some
	  $t \in [j-1]$.  Let $\mathbf{S}$ be the state vector of length
	  $\textrm{ServCount}$ specified in \ref{rmk7}, and
	  $T-\{u\}=[\nu]-\{\Pi(1),\dots,\Pi({j-1})\}-\{u\}$.  By the definition
	  of decodable set function, there exists a state $\mathbf{S'}$ that
	  decodes $\Pi(t)$ from the first $c$ servers, \begin{align}
		  \label{eq0435}
		  \psi_{\mathbf{S}'}^{([c])}(\varphi_{\mathbf{S}^{'}(1)}^{(1)}(W_{[\nu]}),
		  \varphi_{\mathbf{S}^{'}(2)}^{(2)}(W_{[\nu]}), \ldots,
		  \varphi_{\mathbf{S}^{'}(c)}^{(c)}(W_{[\nu]}))=W_{\Pi(t)},
	  \end{align} such that \begin{align} \mathbf{S'}(i)\begin{cases}
		  =\mathbf{S}(i), & i \in [k],  \\ \subseteq
		  [\nu]-\{\Pi(1),\dots,\Pi(j-1)\}-\{u\}, & i \in [k+1, c].
	  \end{cases} \label{eq0504} \end{align}
    
  { If $A_t=1$, then by \ref{rmk7}, $\mathbf{S'}(i) \subseteq
  [\nu]-\{\Pi(1),\dots,\Pi(t)\}$ for all $i\in[c]$. By Remark \ref{rmk_decode}
  in Section \ref{sec2}, we know $\psi_{\mathbf{S}'}^{([c])}$ should return a
  value corresponding to a version in $\cup_{i\in[c]}\mathbf{S'}(i) \subseteq
  [\nu]-\{\Pi(1),\dots,\Pi(t)\}$, which contradicts \eqref{eq0435}. So we
  assume $A_t > 1$.  }
  
  Consider the last iteration with $\textrm{ServCount}$ $=A_t-1$. Let
  $\textrm{VerCount}=x$ in this iteration. By \ref{rmk10}, we know that $x\le
  t$. We will show that if (\ref{eq:contradictthis}) holds, then, in this iteration,
  Line \ref{l6} returns true. Therefore, in this iteration, $\textrm{VerCount}$ $\gets
  x+1$ and $\textrm{ServCount}$ $=A_t-1$ remains unchanged. This contradicts our
  assumption that the last iteration with $\textrm{ServCount}$ $=A_t-1$ has
  $\textrm{VerCount}=x$. So, to complete the proof, it suffices to show that
  Line \ref{l6} returns true in this iteration. We show this next.

  For the iteration of the while loop with $\textrm{ServCount}$ $= A_t-1$ and $\textrm{VerCount}=x$,
  \begin{itemize} \item let $\hat{\mathbf{S}}$ be the server states as in \ref{rmk7},
		  \item let
			  $\hat{T}=[\nu]-\{\Pi(1),\dots,\Pi(x-1)\}=\{\Pi(x),\dots,\Pi(\nu)\}$
			  be the set in Line \ref{lT}, \item let $\hat{U} =
				  \left\{W_u: W_u \in
					  \chi_{\textrm{ServCount}|\hat{T}-\{u\}}\left(\hat{\mathbf{S}}(1),\hat{\mathbf{S}}(2),\ldots,\hat{\mathbf{S}}(\textrm{ServCount}),
					  W_{[\nu]}\right), u \in \hat{T}
				  \right\}$ be the set in Line \ref{l5}.

  \end{itemize} {Here note that $\Pi(t) \in \hat{T}$ because $x \le t$ by \ref{rmk10}.} By
  \ref{rmk7} and \ref{rmk11},  and \eqref{eq0504}, we note that \begin{align}
	  \mathbf{S'}(i)\begin{cases} =\mathbf{S}(i)=\hat{\mathbf{S}}(i), & i
		  \in [A_t -1], \\ \subseteq \hat{T}-\Pi(t), & i \in [A_t,c].
	  \end{cases}\label{eq0410} \end{align} Combining \eqref{eq0410} and
	  \eqref{eq0435}, we know that $\Pi(t)$ is in the decodable set
	  function at $\textrm{VerCount}$ $=t$ and $\textrm{ServCount}$ $=A_t-1$ using the state
	  $\mathbf{S}'$, that is, $$W_{\Pi(t)} \in
	  \chi_{A_t-1|\hat{T}-\{\Pi(t)\}}\left(\mathbf{S}(1),\mathbf{S}(2),\ldots,\mathbf{S}(A_t-1),
	  W_{[\nu]}\right),$$ which combined with the fact that $\Pi(t) \in
	  \hat{T}$ implies $\hat{U} \neq \emptyset$ and Line \ref{l6} returns
	  true. Thus $\textrm{VerCount}$ $\gets x+1$ and $\textrm{ServCount}$ $=A_t-1$ should stay
	  unchanged. 
	  Hence we get a contradiction.
  
  {(ii) Consider the iteration when $\Pi(\textrm{VerCount})$ is set in Line
  \ref{lpi}, which has $\textrm{VerCount}$ $=t$, $\textrm{ServCount}$ $=A_t$. After Line \ref{lT} is
  executed, we have $T = \{\Pi(t), \Pi(t+1), \ldots, \Pi(\nu)\}$. We know Line
  \ref{l6} returns true, and by Line \ref{lW}, $W_{\Pi(t} =
  \max U$. Thus letting $u=\Pi(t) \in T$, we have $$W_u \in
  \chi_{A_t|T-\{u\}}\left(\mathbf{S}(1),\mathbf{S}(2),\ldots,\mathbf{S}(A_t),
  W_{[\nu]}\right).$$ Moreover, $W_u \notin \{W_{i}: i \in T-\{u\}\}$. Combined
  with statement (i), we obtain the desired statement.} \end{IEEEproof}

\begin{lemma}\label{lem0429} In any execution of Algorithm \ref{fig_alg}, at
	the beginning of the final iteration of the while loop, we have
	$\textrm{VerCount}$$=\nu$, {and Line \ref{l6} returns true in that iteration}.
\end{lemma} \begin{IEEEproof} We prove the lemma by contradiction. Suppose
	that, in an execution of Algorithm \ref{fig_alg}, the last iteration
	begins with $\textrm{VerCount}=j$, for some $j < \nu$. This means we
	have found a set of versions in the set $\textrm{VersionsEncountered}$
	such that Line \ref{l6} is not satisfied for $\textrm{VerCount}=j$.
	Furthermore, knowing
	$\textrm{VersionsEncountered}=\{\Pi(1),\dots,\Pi(j-1)\}$, we note that
	Line \ref{l6} is satisfied for version $\Pi(i),$ when
	$\textrm{ServCount}$ is equal to $A_{i}$ for $i \in [j-1]$. 

 Consider the last iteration of the while loop, $\textrm{VerCount}=j,
 \textrm{ServCount}=c$. 
  The states of the first $c$ servers are shown in \ref{rmk7}.  Note that there
  exists a latest common version among these $c$ server states, which is $\max
  ([\nu]-\{\Pi(1),\dots,\Pi(j-1)\})$. Therefore, the decoding function
  $\psi_{\mathbf{S}}^{([c])}$ returns a non-null value. Specifically, for some
  $u \in [\nu]$, we have
  $$\psi_{\mathbf{S}}^{([c])}(\varphi_{\mathbf{S}(1)}^{(1)}(W_{[\nu]}),
  \varphi_{\mathbf{S}(2)}^{(2)}(W_{[\nu]}), \ldots,
  \varphi_{\mathbf{S}(c)}^{(c)}(W_{[\nu]}))=W_u.$$ Furthermore, because
  $\textrm{ServCount}=c,$ we have \begin{align*} &
	  \chi_{\textrm{ServCount}|T-\{u\}}\left(\mathbf{S}(1),\mathbf{S}(2),\ldots,\mathbf{S}(\textrm{ServCount}),
	  W_{[\nu]}\right) \\ &= \{
		  \psi_{\mathbf{S}}(\varphi_{\mathbf{S}(1)}^{(1)}(W_{[\nu]}),
		  \varphi_{\mathbf{S}(2)}^{(2)}(W_{[\nu]}), \ldots,
		  \varphi_{\mathbf{S}(c)}^{(c)}(W_{[\nu]}))\} = \{W_{u}\}
	  \end{align*} {The above result combined with statement (i) of Lemma \ref{lem0428}
  implies that} the set $U \neq \emptyset$ and Line \ref{l6} is satisfied,
  which contradicts our assumption. This completes the proof.  \end{IEEEproof}

\subsection{Proof of Theorem \ref{thm:converse3}} \label{secB}

Now we are ready to prove the lower bound in Theorem \ref{thm:converse3}.
  
{\begin{IEEEproof}[Proof of Theorem \ref{thm:converse3} for general $\nu$]
	Suppose there is a $(c,c,\nu,M,q)$ multi-version code. Run Algorithm
	\ref{fig_alg} on every $\nu$-tuple distinct version values $W_{[\nu]}$.
	By Lemma \ref{lem0429}, we know the algorithm terminates with
	$\textrm{VerCount}$ $=\nu$ and $1 \le A_1 \le \dots \le A_{\nu} \le c$. We use a
	dummy variable $A_0=1$. First, since the algorithm is deterministic, we
	know there is a mapping $\textbf{AuxVars}$ from
	$W_{[\nu]} \in \mathcal{W}$ to $Y_{[c-1]},Z_{[\nu]},A_{[\nu]},\Pi$.
	Next, we show that $\textbf{AuxVars}$ is a one-to-one
	mapping, that is, we create a mapping from
	$Y_{[c-1]},Z_{[\nu]},A_{[\nu]},\Pi$ to $W_{[\nu]} \in \mathcal{W}$.
 
 We now describe how to obtain $W_{[\nu]} \in \mathcal{W}$ from the output of
 $\textbf{AuxVars}$. In particular, for any $t \in \{1,2,\ldots,
 \nu\}$, we describe a procedure to obtain $W_{\Pi(t)}$ from
 $Y_{1},Y_{2},\ldots,Y_{A_{t}},$ $Z_{t}$ and $W_{\Pi(t+1)}, W_{\Pi(t+2)},$ 
 $\ldots, W_{\Pi(\nu)}$. The procedure automatically implies that we can obtain
 $W_{[\nu]}$ from $Y_{[c-1]}, A_{[\nu]},Z_{[\nu]}$ and $\Pi$.  For any
 realization of distinct values $W_{[\nu]} \in \mathcal{W}$, if we are given
 $\Pi,A_{[\nu]}$, we can set the state to be 
 $$\mathbf{S}(i)= \begin{cases}
	 [\nu]-\{\Pi({1}),\dots,\Pi(j-1)\}, & i\in [A_{j-1},A_j-1], \forall j
	 \in [t], \\ [\nu]-\{\Pi({1}),\dots,\Pi(t-1)\}, & i = A_t.
 \end{cases}$$ 
 By \ref{rmk7}, the above states are the same as the states in
 Algorithm \ref{fig_alg} in iteration of the while loop with $\textrm{VerCount}$ $=t$,
 $\textrm{ServCount}$ $=A_t$. Note that at that iteration, by \ref{rmkYZ} we know $Y_{[A_t-1]},Z_{t}$ are the
 values of Servers $[A_t]$, which corresponds to the above states
 $\mathbf{S}(1),\dots,\mathbf{S}(A_t)$. That is $Y_{i} =
 \varphi_{\mathbf{S}(i)}(W_{[\nu]}),$ for $i \in [A_t-1]$ and $Z_{t} =
 \varphi_{\mathbf{S}(A_t)}(W_{[\nu]}).$ 
 Let $T = [\nu]-\{\Pi(1),\dots,\Pi(t-1)\}$.  Thus, 
 $W_{T-\{\Pi({t})\}}= W_{\{\Pi(t+1),\dots,\Pi(\nu)\}}$. From Lemma
 \ref{lem0428} (ii), we know that
 $$\{W_{\Pi(t)}\}=\chi_{A_t|T-\{\Pi(t)\}}\left(\mathbf{S}(1),\mathbf{S}(2),\ldots,\mathbf{S}(A_t),
 W_{[\nu]}\right)-\{W_i, i \in T-\Pi(t)\}  .$$ Therefore, to obtain
 $W_{\Pi(t)},$ it suffices to evaluate the set $$
 \chi_{A_t|T-\{\Pi(t)\}}\left(\mathbf{S}(1),\mathbf{S}(2),\ldots,\mathbf{S}(A_t),
 W_{[\nu]}\right)-\{W_i, i \in T-\Pi(t)\} .$$ \ref{rmk9} states that the above
 set can be computed using $Y_{[A_t-1]},Z_{t}, W_{T-\{\Pi(t)\}}$ via equation (\ref{eq:rmk9}).
 Therefore,  we can compute $W_{\Pi(t)}$ as \begin{align*}& \{ W_{\Pi(t)} \}
	 =\\ &= \bigg\{\psi_{\mathbf{S'}}^{([c])}(X_1,X_2 \ldots, {X}_{c}) \neq
	 \textrm{NULL}: \nonumber \\
	 &\mathbf{S'}=(\mathbf{S}(1),\ldots,\mathbf{S}({A_t}),
	 {S'}_{A_{t}+1},\dots,{S'}_c), \forall {S'}_j\subseteq T-\{\Pi(t)\}, j \in
	 [l+1,c], \nonumber \\ & X_m = Y_m, 1 \leq m \leq A_t-1  \nonumber \\ &
	 X_m = Z_t, m = A_t, \nonumber \\ & X_m = \varphi_{{S'}_m}(W_{S'_m}),
	 A_t+1 \leq m \leq c \bigg\} \\
	 &- \{W_{i}: i \in
	 T-\{\Pi(t)\}\}.   \end{align*} Therefore, given $
	 W_{\{\Pi(t+1),\dots,\Pi(\nu)\}}$, $Y_{[A_t-1]},Z_{t}$,  $A_{[\nu]}$,
	 $\Pi$,  we see that the value of $W_{\Pi(t)}$ is determined.
  {Noting that $A_i \le c$ for all $i \in [\nu]$ by Lemma \ref{lem0429}, we
  infer that, given $Y_{[c-1]},Z_{[\nu]},A_{[\nu]},\Pi$, we can determine
  values of $W_{\Pi(\nu)},W_{\Pi(\nu-1)},\dots,W_{\Pi(1)}$ one by one. Hence we
  have a mapping from $Y_{[c-1]},Z_{[\nu]},A_{[\nu]},\Pi$ to $W_{[\nu]} \in
  \mathcal{W}$, implying that $\textbf{AuxVars}$ is one-to-one.}
  Then we know \eqref{eq:converseequation} and hence \eqref{eq0507} are
  satisfied,  thus the theorem is proved.
%
\end{IEEEproof}}

\begin{remark}
	If $W_{[\nu]}$ is not uniformly distributed, then the proof of Theorem \ref{thm:converse3} can be appropriately modified 
to obtain a lower bound on the storage size per server using \eqref{eq0507}: $$\log q
\ge \frac{ H(W_{[\nu]}|\mathbbm{1}_{W_{[\nu]} \in \mathcal{W}}=1) -
\log(\nu!\binom{c+\nu-1}{\nu})}{c+\nu}.$$ The above bound can be much smaller
compared to the one in Theorem \ref{thm:converse3}, if, for example, the versions are
dependent and have some structure. In particular, the storage cost lower bound would be much smaller if, because of the dependency of the versions, $H(W_{[\nu]}|\mathbbm{1}_{W_{[\nu]} \in
\mathcal{W}=1})) << \log |\mathcal{W}|.$ In this case, it is an open problem
to study codes that exploit the dependency amongst versions to obtain a smaller storage cost.
\label{remark:dependent}
\end{remark}

\section{Toy Model of Distributed Storage}
\label{sec:toy_example}

\begin{figure}
  \centering
  \includegraphics[width=0.48\textwidth]{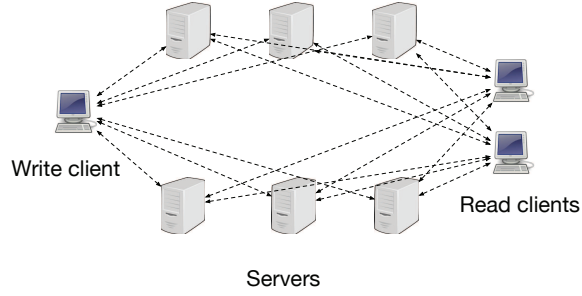}\\
  \caption{System architecture of a toy model with one write client and many read clients.}\label{fig:System_Architecture}
\end{figure}

The multi-version coding problem has no temporal aspect in its formulation. Here, we study a toy model of storage system that evlolves over time and demonstrate the potential of multi-version codes for an asynchronous setting. In particular, we explicitly describe an arrival model for new versions and channel models for the links between the encoders, servers and decoders. Our study of the toy model establishes a physical interpretation for the parameter $\nu$ of multi-version coding. In particular, our toy model demonstrates the connection between the parameter $\nu$ and the degree of asynchrony in a storage system. We begin with a description of our model. 

Consider a distributed storage system with $N$ servers, a write client that generantes different versions of the message, and read clients that aim to to read the message versions (See Fig. \ref{fig:System_Architecture}). The write client aims to store the new version of the message in a distributed storage system consisting of the servers. We aim to design server storage strategies that implement a consistent distributed storage system, and are tolerant to $f$ server failures. We now describe toy models for the channels between the clients and the servers, and the arrival model at the clients. 

\textbf{Message arrival model:} In the toy model here, we assume that a new version of the message appears at the write client in every time-slot. The message at time slot $t$ is denoted as $W_t$, $t \in \mathbb{N}^+,$ where $W_{t} \in [M]$. 
At time slot $t$, the write client sends a packet containing time stamp and the message, that is a packet with $(t,W_t),$ to every server.

\textbf{Channel Model:}
We describe two models for the channels between the clients and the servers. The first model is a delay based model, and the second model is an erasure based model. 

\emph{Delay model:} We assume that a transmitted packet sent by the write client at time slot $t$ to a server arrives at the server in any one of the time slots $\{t,t+1,\ldots,t+T-1\}.$ In other words, the sent message has a delay that can be any one of $0,1,2,\ldots,T-1$. The delay is not known a priori, and can be different for different packets. It is useful to note that even if the same packet is sent in the same time slot to different servers, the delay can be different for different packets.

Note that in the message arrival model, for every time slot $t$, there is a message $(t,W_t)$ sent to every server. Let $\mathcal{S}_{m}^{(t)} \subset \{1,2,\ldots, t\}$ denote the set of versions received by server $m$ at time $t$. Then in the delay model 
\begin{align} 
	\mathcal{S}_{m}^{(t)}=\left(\{t-T+1\} \cup R_{m}^{(t)} \right)  - \bigcup_{j=1}^{t-1} \mathcal{S}_{m}^{(j)} , \label{eq:delaymodel}
\end{align}
where $R_{m}^{(t)}$ is some arbitrary subset of $\{t-T+2,\dots,t\}$. The arrival time set $\mathcal{S}_{m}^{(t)}$ at the servers are not known apriori. Our model is adversarial, that is, we want our decoding constraints (specified below) to be satisfied for all possible arrival time sets $\mathcal{S}_{m}^{(t)}$ that are of the form (\ref{eq:delaymodel}).

\emph{Erasure Model:}
In the erasure model for the channel, a packet sent by the write client to the server may be erased. There is no packet delay in the erasure model. Our erasure model is adversarial, with the following packet delivery guarantee: for every subset of $N-f$ servers, for any consecutive $T$ packets, there is at least one least one packet such that it arrives at all the $N-f$ servers. Mathematically, the received packet versions $\mathcal{S}_{m}^{(t)}$ at server $m$ at time $t$ is 

\begin{equation}
\mathcal{S}_{m}^{(t)} = \left\{\begin{array}{cc} \{t\} &\textrm{ if } \exists n_1, n_2,\ldots,n_{N-f-1} \in [n]\textrm{ s.t. } ~~\bigcup_{j=t-T+1}^{t-1} \mathcal{S}_{m}^{(j)} \cap \bigcap_{k=1}^{N-f-1} \mathcal{S}_{n_k}^{(j)} = \phi \\ \{t\} \textrm{ or } \phi &\textrm{otherwise} \end{array}\right.
	\label{eq:erasuremodel}
\end{equation}


\textbf{Encoding requirements:}
At time $t$, for every $m \in [N],$ server $m$ stores a symbol $X_m^{(t)}$  which is a function of the stored symbol $X^{(t-1)}_{m}$ and the received packets $W_{\mathcal{S}_{m}^{(t)}}$. We assume that $X^{(t)}_{m} \in [q]$ for every value of $t,m$. As usual, for a given encoding scheme, we measure its storage cost as $\frac{\log q}{\log M}$. 

\textbf{Decoding requirements:}
We intend to design a failure tolerance of $f$ servers. In our decoding requirement, a read client accesses any subset of $N-f$ servers and requires to decode the latest common message version among the $N-f$ servers, or the message corresponding to a later version. Our model is adversarial, that is, in the delay model, we want the read client decode the latest common verison for every possible packet arrival pattern at the servers that satisfies (\ref{eq:delaymodel}). Similarly, in the erasure model, we intend the read client to decode the latest common version for every possible packet arrival pattern that satisfies (\ref{eq:erasuremodel}).   In our toy model we assume that the link between the servers and read clients are perfect, that is there is no erasure or delay for the packets between the servers and the read client.

It is instructive to note that both the erasure model and delay model ensure that there is at time $t$, there is a latest common version for all the servers among the versions $\{t, t-1, \ldots, t-T+1\}$. Therefore, under our decoding requirements, a read client that aims to read from the storage system at time slot $t$ must decode a message in $\{W_{t-T+1}, W_{t-T+2},\ldots,W_{t}\}$. 

We will see that the multi-version coding problem can be used to reveal the fundamental storage cost performance of this setting. In particular, our achievability and converse results of an $(n,c,\nu)$ multi-version code provides insights on the storage cost in our setting, where $n=N, c=N-f$ and $T=\nu$. 

\begin{claim}
	Consider a setting where very new message version at the write client takes values from the set $[M]$. If $T | (N-f-1),$ then a storage cost of $$\log q=\frac{T}{T+N-f-1}\log M + o (\log M)$$ is achievable. 
\label{claim10828}
\end{claim}
\begin{IEEEproof}
	The proof is the same for both the erasure model and the delay model. Observe that, in both models, at time $t$, there is a latest common version among all the servers in $[t-T+1,t]$. As per Construction \ref{cnstr141001} for an $(N,T,N-f)$ multi-version code, each server stores $\frac{\log M}{T+N-f-1}$ bits of the latest version it has received. Along the same lines as the proof of Theorem \ref{thm:3}, we infer that any read client that connects to $N-f$ servers at time $t$ gets at least $\lceil \frac{N-f}{T}\rceil = \frac{N-f+T-1}{T}$ codeword symbols corresponding to at least one version ${\nu}^*$, where $\nu^{*} \in [t-T+1, t]$ is the latest common version or a later version among the $N-f$ servers. Therefore, version $\nu^{*}$ is decodable by the read client which reads at time $t$. There is a a storage overhead of $o(\log_2 M)$ bits since the servers need to store the \emph{time stamp}\footnote{In fact, a server that stores a codeword sybol corresponding to message $W_{t}$ can store the time stamp as $t \mod 2T.$ We omit mechanical details of the proof here.} of the version along with the codeword symbol.
\end{IEEEproof}

\begin{claim}
 Consider a setting where very new message version at the write client takes values from the set $[M]$. The storage cost of any server storage strategy for the delay model satisfies
$$\frac{\log q}{\log M} \ge
	\frac{T-1}{N-f+T-2}-\frac{\log((T-1) ^{T-1}
	\binom{N-f+T-2}{T-1})}{(N-f+T-2) \log M }.$$
\label{claim2}
\end{claim}

\begin{claim}
Consider a setting where very new message version at the write client takes values from the set $[M]$. The storage cost of any server storage strategy for the erasure model satisfies
$$\frac{\log q}{\log M} \ge
	\frac{T}{N-f+T-1}-\frac{\log(T ^{T}
	\binom{N-f+T-1}{T})}{(N-f+T-1) \log M }.$$
\label{claim3}
\end{claim}

Claims \ref{claim2} and \ref{claim3} are essentially corollaries to Theorem \ref{thm:converse3}. In particular, for both the delay model and the erasure model, we note that the server encoding functions necessarily implements a multi-version code. We provide brief sketches of their proofs here.
\begin{IEEEproof}[Proof of Claim \ref{claim2}]
Consider an arbitrary collection of subset $\mathcal{A}_{1},\mathcal{A}_{2},\ldots,\mathcal{A}_{n} \subseteq [t-T+2,t]$. 
Assume that at time $t-T,$ all the packets from the write client to the server
are delivered. At times $t-T+1,\ldots, t-1,$ no packet is delivered by the
channel. At time $t,$ server $m$ receives packets $\mathcal{A}_{m} \cup
\{t-T+1\}$. Given the versions $1,2,\ldots,t-T+1$, the server encoding
functions at time $t$ form a multi-version code over the versions in
$[t-T+2,t].$ Since we started with an arbitrary collection of subsets $\mathcal{A}_{1},\mathcal{A}_{2},\ldots,\mathcal{A}_{n}$, the worst storage cost over all collections of subsets is lower bounded by the cost described in Theorem \ref{thm:converse3} for $\nu=T, c=N-f$.  This completes the proof.
\end{IEEEproof}

\begin{remark}
	The worst case states described the converse of Theorem \ref{thm:converse3}, when applied to the delay model, may implicitly require the packets sent to be delivered out of order. This is because, when the converse of Theorem \ref{thm:converse3} is applied in our proof of Claim \ref{claim2}, we may require a server $m$ to contain a version $t_1$ but not contain a version $t_2,$ where $t_2 < t_1, t_2, t_1 \in [t-T+2,t]$. Our converse for the delay model is therefore  more relevant to applications where the versions may be sent in one order, and received in another. The assumption that the order of packets may change is the basis of certain transport protocols such as Stream Control Transmission Protocol (SCTP) \cite{stewart2007stream}.

\end{remark}

\begin{IEEEproof}[Proof of Claim \ref{claim3}]
Consider an arbitrary collection of subset $\mathcal{A}_{1},\mathcal{A}_{2},\ldots,\mathcal{A}_{n} \subseteq [t-T+1,t]$ such that, for every collection of $c$ subsets in $\mathcal{A}_{1},\mathcal{A}_{2},\ldots,\mathcal{A}_{n},$ there is a common version. Assume that at time $t_0$ in $[t-T+1,t]$, server $m \in [n]$ receives the packet sent by the write client if and only if $t_0 \in \mathcal{A}_{m}$. Given the messages $W_{[t-T]},$ the server encoding strategy at times $[t-T+1:t]$ forms a multi-version code. Since we started with an arbitrary collection of subsets $\mathcal{A}_{1},\mathcal{A}_{2},\ldots,\mathcal{A}_{n}$, the worst storage cost over all collections of subsets is lower bounded by the cost described in Theorem \ref{thm:converse3} for $\nu=T, c=N-f$. This completes the proof.
\end{IEEEproof}

The parameter $\nu$ is analogous to the parameter $T$ in both the erasure and the delay models. The parameter $T$ is, intuitively speaking, a measure of the degree of asynchrony in the system. Our toy models therefore establishes an explicit connection between the parameter $\nu$ and the degree of asynchrony in the storage system. A multi-version code with a larger value for the parameter $\nu$ can tolerate a greater degree of asynchrony, albeit at a larger storage cost.

\section{Concluding Remarks}\label{sec:conclude} In this paper, we have
proposed the multi-version coding problem, where the goal is to encode various versions in a distributed storage system so that the latest version is decodable. 
We have given a lower bound on
the worst-case storage cost and provide a simple coding scheme that is
essentially optimal for an infinite family of parameters. Our problem formulation and solution is a step towards the study of consistent key value stores from an information theoretic perspective. The multi-version coding problem affords a number of interesting generalizations which are relevant to practical consistent distributed storage systems.  We discuss some of these generalizations next.

\begin{itemize}

	\item A useful direction of future work is to study the
problem beyond a worst-case setting, for instance, through analysing a restricted set of states. For example, one can assume that the
servers always get consecutive versions: $\mathbf{S}(i) = [x,y]$, for some $1
\le x \le y \le \nu$. One can similarly assume that due to network constraints,
certain versions only are dispersed to a subset of the servers, namely,  $x
\notin \mathbf{S}(i)$ for all $i \in I$, where $I \subseteq [n]$ is some subset
of server indices. More generally, our problem could be
formulated in terms of storage cost per server per state, and the overall storage cost can be optimized based on the workload distributions of the servers.

\item Our problem formulation assumes that the number of versions $\nu$, is known a priori. An interesting direction of future work is to manipulate our problem formulation and solutions to incorporate a setting where this parameter is not known. 
\item Our problem formulation essentially views different versions as being independent. However, in several applications, it is conceivable that different versions are correlated. For dependent versions, Remark \ref{remark:dependent} suggests that the converse of Theorem \ref{thm:converse3} would be applicable after appropriate manipulations. Developing code constructions that exploit dependency in the versions is an interesting area of future work. The ideas of \cite{Mazumdar_update,Oggier_update} can be useful in this endeavor.
\item The framework of our toy model can be developed to study more realistic scenarios. The first step would be to incorporate asynchrony/erasures in the read client. The end goal of the framework would be to understand costs in realistic storage systems, or over models studied in distributed algorithms literature. The standard model in distributed algorithms can be viewed as the delay model of Section \ref{sec:toy_example} in the limiting case of asymptotically large $T$. Furthermore, in distributed computing theory, write and read clients, and servers are modeled as automota (more precisely, input-output automata \cite{Lynch1996,Lynch_tuttle}), the goal is to design client and server protocols that ensure consistency. Developments of our toy model, and appropriate refinements to multi-version coding, can potentially provide information theoretic insights into the storage cost of such systems. 

\item Minimizing communication costs and latency are important requirements of modern
consistent storage services. Refinements of multi-version coding, and our toy model for channels to incorporate these requirements is an important direction of future work. In particular, the tools used in  references \cite{Mazumdar_update, Rouayheb_Synchronizing, Oggier_update,Rawat_update}, when appropriately adopted to multi-version coding, may help reduce latency by reducing the amount of information transmitted to disperse and update information related to a new version. 

\end{itemize}


\section*{Acknowledgment}
The authors would like to thank Prof. Nancy Lynch, Prof. Muriel M\'{e}dard, and Prof. Tsachy Weissman
for their valuable advice and helpful comments. 

This work is partially supported by the Center for Science of Information (CSoI), an NSF Science and Technology Center, under grant agreement CCF-0939370. 

\bibliographystyle{IEEEtran}
\bibliography{mybib}

\begin{thebibliography}{10}
\providecommand{\url}[1]{#1}
\csname url@samestyle\endcsname
\providecommand{\newblock}{\relax}
\providecommand{\bibinfo}[2]{#2}
\providecommand{\BIBentrySTDinterwordspacing}{\spaceskip=0pt\relax}
\providecommand{\BIBentryALTinterwordstretchfactor}{4}
\providecommand{\BIBentryALTinterwordspacing}{\spaceskip=\fontdimen2\font plus
\BIBentryALTinterwordstretchfactor\fontdimen3\font minus
  \fontdimen4\font\relax}
\providecommand{\BIBforeignlanguage}[2]{{%
\expandafter\ifx\csname l@#1\endcsname\relax
\typeout{** WARNING: IEEEtran.bst: No hyphenation pattern has been}%
\typeout{** loaded for the language `#1'. Using the pattern for}%
\typeout{** the default language instead.}%
\else
\language=\csname l@#1\endcsname
\fi
#2}}
\providecommand{\BIBdecl}{\relax}
\BIBdecl

\bibitem{Lynch1996}
N.~A. Lynch, \emph{Distributed Algorithms}.\hskip 1em plus 0.5em minus
  0.4em\relax San Francisco, CA, USA: Morgan Kaufmann Publishers Inc., 1996.

\bibitem{lamport1979make}
L.~Lamport, ``How to make a multiprocessor computer that correctly executes
  multiprocess programs,'' \emph{Computers, IEEE Transactions on}, vol. 100,
  no.~9, pp. 690--691, 1979.

\bibitem{vogels2008eventually}
W.~Vogels, ``Eventually consistent,'' \emph{Queue}, vol.~6, no.~6, pp. 14--19,
  2008.

\bibitem{ABD}
H.~Attiya, A.~Bar-Noy, and D.~Dolev, ``Sharing memory robustly in
  message-passing systems,'' \emph{J. ACM}, vol.~42, no.~1, pp. 124--142, Jan.
  1995.

\bibitem{Decandia}
G.~DeCandia, D.~Hastorun, M.~Jampani, G.~Kakulapati, A.~Lakshman, A.~Pilchin,
  S.~Sivasubramanian, P.~Vosshall, and W.~Vogels, ``{Dynamo: Amazon's highly
  available key-value store},'' in \emph{SOSP}, vol.~7, 2007, pp. 205--220.

\bibitem{CassandradB}
E.~Hewitt, \emph{Cassandra: the definitive guide}.\hskip 1em plus 0.5em minus
  0.4em\relax " O'Reilly Media, Inc.", 2010.

\bibitem{CouchDB}
J.~C. Anderson, J.~Lehnardt, and N.~Slater, \emph{CouchDB: the definitive
  guide}.\hskip 1em plus 0.5em minus 0.4em\relax O'Reilly Media, Inc., 2010.

\bibitem{Herlihy_Shavit}
M.~Herlihy and N.~Shavit, \emph{The Art of Multiprocessor Programming, Revised
  Reprint}.\hskip 1em plus 0.5em minus 0.4em\relax Elsevier, 2012.

\bibitem{Zookeeper}
P.~Hunt, M.~Konar, F.~P. Junqueira, and B.~Reed, ``Zookeeper: Wait-free
  coordination for internet-scale systems.'' in \emph{USENIX Annual Technical
  Conference}, vol.~8, 2010, p.~9.

\bibitem{Hendricks}
J.~Hendricks, G.~R. Ganger, and M.~K. Reiter, ``Low-overhead byzantine
  fault-tolerant storage,'' \emph{ACM SIGOPS Operating Systems Review},
  vol.~41, no.~6, pp. 73--86, 2007.

\bibitem{Dutta}
P.~Dutta, R.~Guerraoui, and R.~R. Levy, ``Optimistic erasure-coded distributed
  storage,'' in \emph{Distributed Computing}.\hskip 1em plus 0.5em minus
  0.4em\relax Springer, 2008, pp. 182--196.

\bibitem{CadambeCoded_NCA}
V.~R. Cadambe, N.~Lynch, M.~Medard, and P.~Musial, ``A coded shared atomic
  memory algorithm for message passing architectures,'' in \emph{2014 IEEE 13th
  International Symposium on Network Computing and Applications (NCA)}.\hskip
  1em plus 0.5em minus 0.4em\relax IEEE, 2014, pp. 253--260, extended version
  available at http://arxiv.org/abs/1407.4167.

\bibitem{Dobre}
D.~Dobre, G.~Karame, W.~Li, M.~Majuntke, N.~Suri, and M.~Vukoli{\'c},
  ``{PoWerStore: proofs of writing for efficient and robust storage},'' in
  \emph{Proceedings of the 2013 ACM SIGSAC conference on Computer \&
  communications security}.\hskip 1em plus 0.5em minus 0.4em\relax ACM, 2013,
  pp. 285--298.

\bibitem{Mazumdar_update}
A.~Mazumdar, G.~W. Wornell, and V.~Chandar, ``Update efficient codes for error
  correction,'' in \emph{Information Theory Proceedings (ISIT), 2012 IEEE
  International Symposium on}.\hskip 1em plus 0.5em minus 0.4em\relax IEEE,
  2012, pp. 1558--1562.

\bibitem{Rouayheb_Synchronizing}
S.~E. Rouayheb, S.~Goparaju, H.~M. Kiah, and O.~Milenkovic, ``Synchronizing
  edits in distributed storage networks,'' \emph{arXiv preprint
  arXiv:1409.1551}, 2014.

\bibitem{Oggier_update}
J.~Harshan, A.~Datta, and F.~E. Oggier, ``Compressed differential erasure codes
  for efficient archival of versioned data,'' \emph{arxiv preprint}, 2015,
  http://arxiv.org/abs/1503.05434.

\bibitem{Rawat_update}
A.~S. Rawat, S.~Vishwanath, A.~Bhowmick, and E.~Soljanin, ``Update efficient
  codes for distributed storage,'' in \emph{2011 IEEE International Symposium
  on Information Theory Proceedings (ISIT)}.\hskip 1em plus 0.5em minus
  0.4em\relax IEEE, 2011, pp. 1457--1461.

\bibitem{Wang-Cadambe-ISIT}
Z.~Wang and V.~Cadambe, ``Multi-version coding in distributed storage,'' in
  \emph{2014 IEEE International Symposium on Information Theory (ISIT)}.\hskip
  1em plus 0.5em minus 0.4em\relax IEEE, 2014, pp. 871--875.

\bibitem{Tian_433}
C.~Tian, ``Characterizing the rate region of the (4, 3, 3) exact-repair
  regenerating codes,'' \emph{Selected Areas in Communications, IEEE Journal
  on}, vol.~32, no.~5, pp. 967--975, 2014.

\bibitem{Fragouli_pliable}
S.~Brahma and C.~Fragouli, ``Pliable index coding,'' in \emph{2012 IEEE
  International Symposium on Information Theory Proceedings (ISIT)}.\hskip 1em
  plus 0.5em minus 0.4em\relax Ieee, 2012, pp. 2251--2255.

\bibitem{Fragouli_Content}
L.~Song and C.~Fragouli, ``Content-type coding,'' \emph{arXiv preprint
  arXiv:1505.03561}, 2015.

\bibitem{griva2009linear}
I.~Griva, S.~G. Nash, and A.~Sofer, \emph{Linear and nonlinear
  optimization}.\hskip 1em plus 0.5em minus 0.4em\relax Siam, 2009.

\bibitem{Wang_Cadambe_Allerton}
Z.~Wang and V.~Cadambe, ``On multi-version coding for distributed storage,'' in
  \emph{52nd Annual Allerton Conference on Communication, Control, and
  Computing (Allerton)}, Sept 2014, pp. 569--575.

\bibitem{stewart2007stream}
R.~Stewart, ``Stream control transmission protocol,'' 2007, {R}FC 4960:
  Available at https://tools.ietf.org/html/rfc4960.

\bibitem{Lynch_tuttle}
N.~A. Lynch and M.~R. Tuttle, ``Hierarchical correctness proofs for distributed
  algorithms,'' in \emph{Proceedings of the sixth annual ACM Symposium on
  Principles of distributed computing}.\hskip 1em plus 0.5em minus 0.4em\relax
  ACM, 1987, pp. 137--151.

\end{thebibliography}

\end{document}